\documentclass[]{interact}

\usepackage[numbers,sort&compress]{natbib}
\bibpunct[, ]{[}{]}{,}{n}{,}{,}
\usepackage[utf8]{inputenc}
\usepackage{graphicx}
\usepackage{color}
\definecolor{orange}{rgb}{0.99,.55,0}

\newcommand{\botwo}{\ensuremath{\textrm{BO}_2}}
\newcommand{\pdcoo}{\ensuremath{\textrm{PdCoO}_2}}
\newcommand{\ptcoo}{\ensuremath{\textrm{PtCoO}_2}}
\newcommand{\pdcro}{\ensuremath{\textrm{PdCrO}_2}}
\newcommand{\cucro}{\ensuremath{\textrm{CuCrO}_2}}
\newcommand{\curho}{\ensuremath{\textrm{CuRhO}_2}}
\newcommand{\pdrho}{\ensuremath{\textrm{PdRhO}_2}}
\newcommand{\agnio}{\ensuremath{\textrm{AgNiO}_2}}
\newcommand{\cufeo}{\ensuremath{\textrm{CuFeO}_2}}

\newcommand{\sovert}[1]{\ensuremath{#1\,\mu\textnormal{V K}^{-2}}}
\newcommand{\seebeck}[1]{\ensuremath{#1\,\mu\textnormal{V K}^{-1}}}
\newcommand{\resist}[1]{\ensuremath{#1\,\mu\Omega\,\textnormal{cm}}}
\newcommand{\Resist}[1]{\ensuremath{#1\,\Omega\,\textnormal{cm}}}

\newcommand{\wkm}[1]{\ensuremath{#1\,\textnormal{W K}^{-1}\textnormal{m}^{-1}}}

\newcommand{\pfac}[1]{\ensuremath{#1\,W\textnormal{m}^{-1}\textnormal{K}^{-2}}}

\makeatletter
\def\NAT@def@citea{\def\@citea{\NAT@separator}}
\makeatother

\theoremstyle{plain}

\theoremstyle{definition}

\theoremstyle{remark}

\begin{document}

\articletype{Review}

\title{Unconventional aspects of electronic transport in delafossite oxides}

\author{
\name{Ramzy Daou\textsuperscript{a}, Raymond Fr\'esard\textsuperscript{a}\thanks{Email: Raymond.Fresard@ensicaen.fr}, Volker Eyert\textsuperscript{a,b}, Sylvie H\'ebert\textsuperscript{a}, and Antoine Maignan\textsuperscript{a}}
\affil{Normandie Univ, ENSICAEN, UNICAEN, CNRS, CRISMAT, 14050 Caen, France; 
\textsuperscript{b}Present address:
Materials Design SARL, 42, Avenue Verdier, 92120 Montrouge, France}
}

\maketitle

\begin{abstract}
The electronic transport properties of the delafossite oxides ABO$_2$ are
usually understood in terms of two well separated entities, namely, the 
triangular A$^+$ and (BO$_2$)$^-$ layers. Here we review several cases 
among this extensive
family of materials where the transport depends on the interlayer coupling and
displays unconventional properties. We review the doped thermoelectrics based
on \curho{} and \cucro{}, which show a high-temperature recovery of
Fermi-liquid transport exponents, as well as the highly anisotropic metals
\pdcoo{}, \ptcoo{} and \pdcro{} where the sheer simplicity of the Fermi
surface leads to unconventional transport. We present some of the theoretical
tools that have been used to investigate these transport properties and review
what can and cannot be learned from the extensive set of electronic structure
calculations that have been performed. 
\end{abstract}


\section{Introduction}
\label{intro}

Transition metal oxides attract a lot of attention due to a great 
variety of physical phenomena, most of which go along with the 
ordering of some microscopic degrees of freedom as a function 
of, {\it e.g.}, temperature, pressure, or doping \cite{Ima98}. Prominent examples 
are the striking metal-insulator transitions in vanadium sesquioxide  
\cite{McW73,Hel01,Lim03,Gry07}, high-$ {\rm T_c} $ superconductivity in the cuprates, 
or the colossal magnetoresistance observed in the manganates 
\cite{Hel93,Tom95,Rav95,Mai95}. Cobaltates have 
aroused much interest due to the occurrence of different spin states 
\cite{Eye04,Fre04,Wu05}. In addition, they are promising materials 
for thermoelectric applications \cite{Ter97,Mas00}. 

Known since 1873, when Friedel discovered the mineral $ {\rm CuFeO_2} $ \cite{Fri73}, 
the delafossites $ {\rm ABO_2} $ continue to generate strong and ever 
increasing interest \cite{Sha71,Tan98,Mar06}, especially 
after Kawazoe {\em et al.}\ showed simultaneous transparency and 
p-type conductivity \cite{Kaw97} in ${\rm CuAlO_2}$. This discovery laid the groundwork for the 
development of transparent optoelectronic devices. Furthermore, the 
quasi two-dimensionality of the lattice and the triangular coordination 
of atoms give rise to exciting physical properties such as strong 
anisotropy of the electrical conductivity and magnetic frustration 
effects. 

\begin{figure}[t]
\centering
\includegraphics[width=0.64\columnwidth,clip]{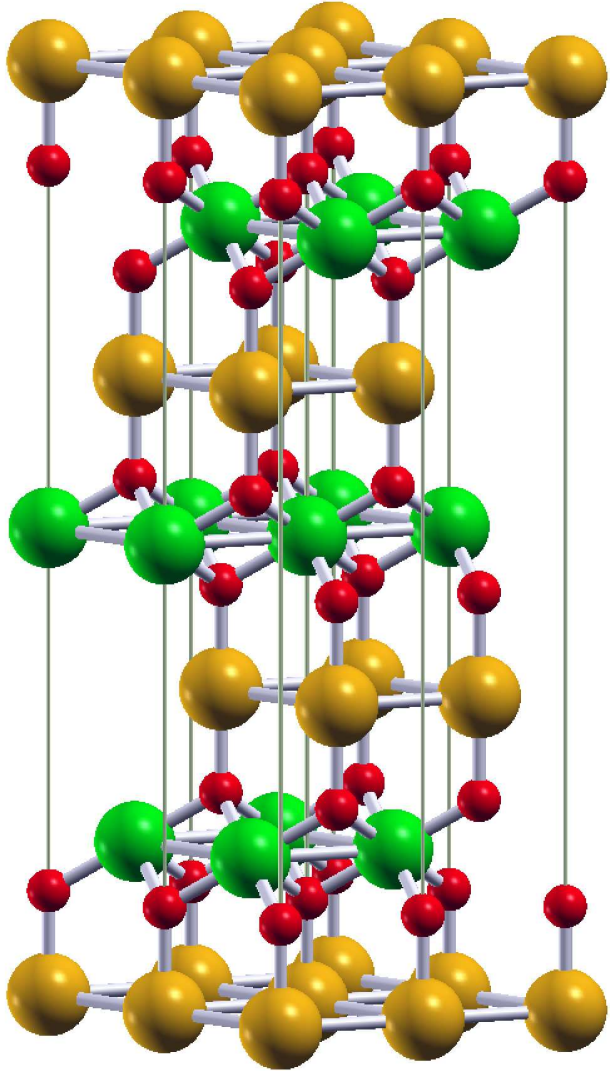}
\caption{Crystal structure of delafossite $ {\rm ABO_2} $. 
         A, B, and oxygen atoms are shown as orange, green, and  red
         (light, big dark, and small dark in grayscale) spheres, respectively.}
\label{figstruct}
\end{figure}

The delafossite structure has the space group $ {\rm R\bar{3}m} $ and 
results from a stacking of monoatomic triangular layers, see
Fig.~\ref{figstruct} \cite{Sha71,Mar06}. 
In particular, the B-atoms are at the centers of edge-sharing distorted oxygen 
octahedra, which form the characteristic $ {\rm BO_2} $ 
sandwich layers. These trilayers are interlinked by linear O--A--O 
bonds, resulting in a twofold coordination of the A-atoms. However, 
the latter have, in addition, six in-plane nearest neighbor A-atoms.  
For this reason, the structure may be likewise regarded as formed 
from single A-atom layers, which are intertwined by the octahedral 
sandwiches. We find this point of view particularly useful 
when investigating the metallic delafossites. Finally, the oxygen 
atoms are tetrahedrally coordinated by one A-atom and three B-atoms. 
Pressure studies on $ {\rm PdCoO_2} $ and $ {\rm PtCoO_2} $ reveal an 
increase of the structural anisotropy on compression indicating the 
high mechanical stability of both the octahedral sandwich layers and 
the O--Pd--O (O--Pt--O) dumbbells \cite{Has03}.  

Generically, the A and B atoms are mono- and trivalent, respectively. 
Depending on the chemical composition, a wide variety of behaviors is
therefore possible.  
For instance, if the $ {\rm A^+} $ ion is in a $ d^9 $ configuration, 
metallic conductivity is observed as in the case of 
$ {\rm PdCoO_2} $. If it is in a $ d^{10} $ configuration, 
the degrees of freedom dominating the low-energy physics can be traced back
to the B atoms as, {\it e.g.}, in $ {\rm CuCrO_2} $, $ {\rm AgNiO_2} $
\cite{Waw07,Kan07} 
and $ {\rm AgCrO_2} $ \cite{Ooh94,Sek08}.  

In general, interest in the delafossite-type compounds has focused 
on the triangular arrangement of the transition-metal atoms 
and the resulting possible frustration effects, which arise once 
localized magnetic moments are established. While most of these oxides 
have been found to be antiferromagnetic semiconductors, other class 
members like $ {\rm PdCrO_2} $, $ {\rm PdCoO_2} $, $ {\rm PdRhO_2} $, and 
$ {\rm PtCoO_2} $ attracted interest due to their rather high metallic 
conductivity. 
In particular, $ {\rm PdCoO_2} $ has been shown to possess one of the lowest 
electric resistivities of normal-state oxides, even lower than that of 
Pd metal at room temperature \cite{Sha71,Tan98,Has02}. In fact, with
$\rho(300~K) \simeq \resist{2.6}$ it is even comparable to pure Au.
Yet, 
the conductivity is strongly anisotropic \cite{Sha71,Has02}.  
In particular, the ratio of the resistivities parallel and perpendicular 
to the $ c $ axis can be $ 400 $ or more  in $ {\rm PdCoO_2} $ 
\cite{Has02, McK2017}. 

Despite their simple chemical formulae the delafossites may be regarded 
as prototypical superlattices where the composition of both the A and 
B layers can be used to strongly influence the behavior of the whole 
system. For instance, in $ {\rm CuCrO_2} $, the Fermi energy falls into 
the Cr $ 3d $ band, but since the Cr layers order magnetically this 
compound is a magnetic semiconductor. In contrast, as will be shown 
below, in $ {\rm PdCoO_2} $, the Co layers only act as charge reservoirs, 
and conduction takes place almost exclusively in the Pd layers. 

The paper is organized as follows: The experimental results on thermoelectric
delafossites CuCr$_{1-x}$Mg$_x$O$_2$ and CuRh$_{1-x}$Mg$_x$O$_2$ are reviewed in
Section~\ref{sec:Sylvie}. The properties of metallic delafossites are reviewed
in Section~\ref{sec:Ramzy}. In Section~\ref{sec:theo_ana} we present the
theoretical analysis of metallic and thermoelectric delafossites that make up
the main body of this review. Conclusions and perspectives are presented in 
Section~\ref{sec:concl}. 

\section{Thermoelectric delafossites: doped \cucro{} and \curho{} }\label{sec:Sylvie}

\begin{figure}[t!]
	\centering
\hspace*{-1em}\includegraphics[width=0.9\columnwidth]{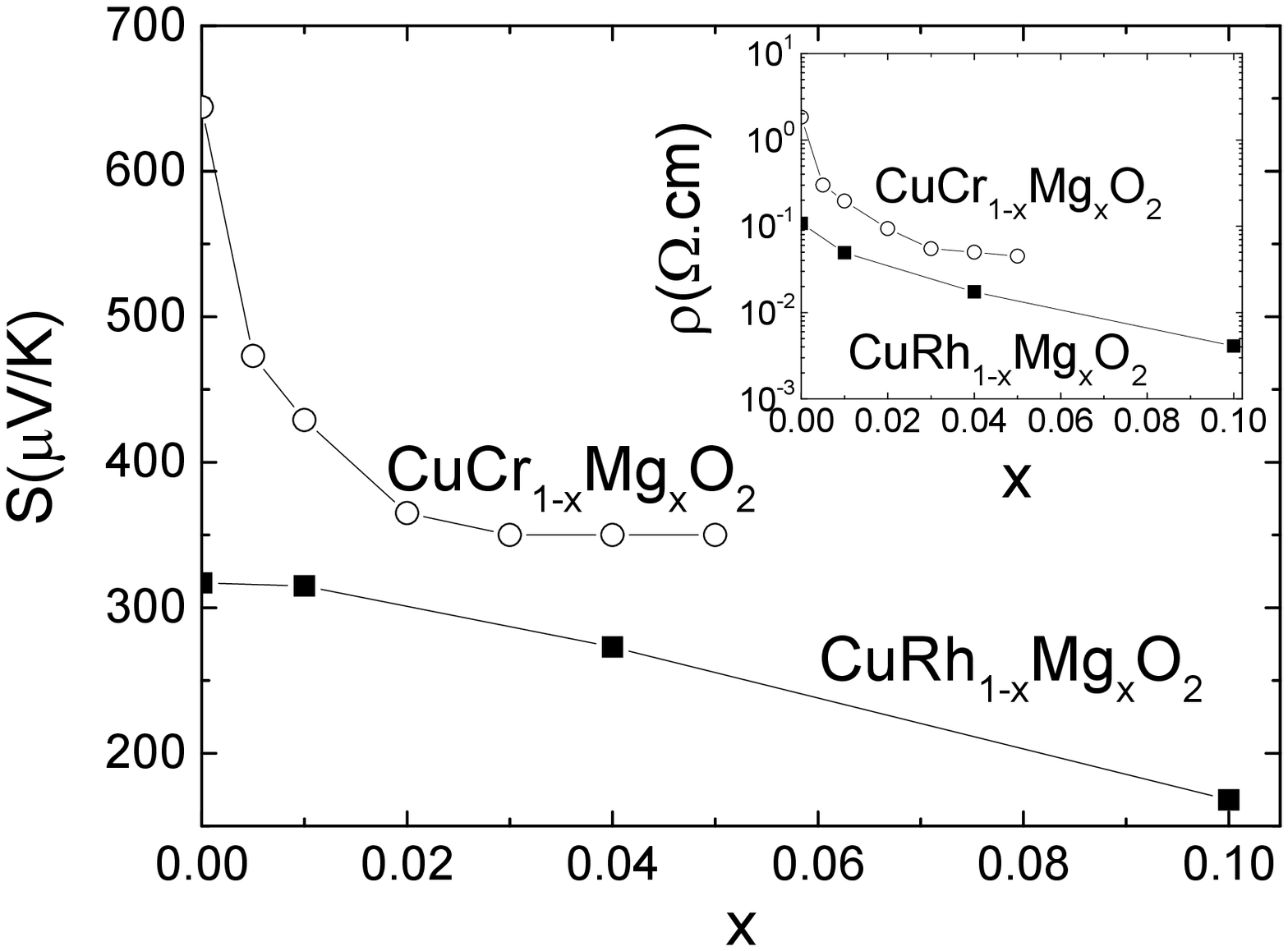}
\caption{Dependence on Mg content $x$ of the thermopower at 300~K of
CuCr$_{1-x}$Mg$_x$O$_2$ and CuRh$_{1-x}$Mg$_x$O$_2$. Inset: Dependence on Mg
content $x$ of the resistivity at 300~K of CuCr$_{1-x}$Mg$_x$O$_2$ and
CuRh$_{1-x}$Mg$_x$O$_2$. 
	}
	\label{fig:Sylvie}
\end{figure}

Many delafossites behave as semiconductors, with a band gap of order 1~eV, and
have been investigated in detail for possible applications in the field 
of transparent conducting oxides (TCO) \cite{Kaw97}. The delafossite family
consists in a large number of materials AMO$_2$, with A = Ag$^+$, Cu$^+$
\ldots~and 
M = Al, Ga, Sc, In, Fe, Cr, Rh \ldots~. In this part, we present the results
that have motivated our investigation of delafossite electronic structure 
and their transport properties. We focus on the results related to
the doping of CuCrO$_2$ and CuRhO$_2$, by substituting Cr$^{3+}$ or
Rh$^{3+}$ by another cation such as Mg$^{2+}$, to increase electrical
conductivity in order to optimize the thermoelectric properties, 
as shown in Fig.~\ref{fig:Sylvie}. The positive Seebeck coefficient for
both pristine delafossites indicates their p-type character and the aliovalent
substitution creating more holes in the 3d or 4d bands explains why $S$ and
$\rho$ decrease with $x$. The properties of metallic 
delafossites with d$^9$ A atom (PdCoO$_2$, PtCoO$_2$, PdCrO$_2$ and PtCrO$_2$)
are described in Section~\ref{sec:Ramzy}. In contrast, CuCrO$_2$ is
semiconducting, with 
a magnetic transition observed at $T_N = 24$~K towards an antiferromagnetic
state. The ratio between $\Theta_p$ and 
$T_N$ is very large, close to 7-8 suggesting a large magnetic frustration
\cite{Dou86,Poi09,Alb17}. The possible
reduction of the gap by doping has stimulated the investigation of transport
properties, and more specifically and more recently, due to the CdI$_2$ type
nature of the CrO$_2$ layers, isostructural to CoO$_2$ layers in Na$_x$CoO$_2$,
thermoelectric properties have been measured. In 2005, a first report was
presented on CuCr$_{1-x}$Mg$_x$O$_2$ \cite{Oku05}, probed by specific heat,
magnetization and transport measurements up to 300~K. By substituting
Mg$^{2+}$ on the Cr$^{3+}$ site, the N\'eel temperature $T_N$ is kept
unchanged as shown by a neutron diffraction study \cite{Poi09}, but
the Curie-Weiss temperature $\Theta_p$ 
increases from $-170$~K to $-100$~K. Considering that Cr$^{4+}$ is 
very difficult to stabilize without high oxygen pressure, the doping effect in
this article was interpreted taking into account a mixed valency of Cu$^+$ /
Cu$^{2+}$ induced by the Mg$^{2+}$ substitution on the Cr$^{3+}$
site. Magnetic susceptibility was analyzed taking into account the
contribution of  both Cr$^{3+}$ ($S = 3/2$) and Cu$^{2+}$ ($S = 1/2$).

There is a direct impact of the antiferromagnetic ordering on the resistivity
curves, with an enhanced magnetoresistance around $T_N$ \cite{Oku05}. The
transport results (magnetoresistance and resistivity) are consistent with a
direct coupling between the doped holes and the Cr$^{3+}$ spin. The
thermopower remains rather large (still $\sim 100 \mu V/K$ for $x =0.02$),
with values close to the ones of Na$_x$CoO$_2$, but resistivity is actually
too high to ensure a large power factor. The authors thus concluded that these
delafossites could not be considered for thermoelectric applications. 

High-temperature properties were investigated by Ono {\em et
al.} \cite{Ono07}, with electrical resistivity, Seebeck coefficient and thermal
conductivity measured up to 1100~K and for $x \leq 0.05$. In this paper,
contrary to the interpretation of Okuda {\em et al.} \cite{Oku05}, the authors argue
that a mixed valency of Cr$^{3+}$ and Cr$^{4+}$ is induced by the Mg$^{2+}$
substitution on the Cr site. The Seebeck coefficient dependence on $x$ could
indeed be interpreted considering the Koshibae and Maekawa formula
\cite{Kos00}. The Cr$^{3+}$ and Cr$^{4+}$ are supposed to be in the high-spin
state ($S = 3/2$ and $S =  1$ respectively), and an extra spin entropy term of
\seebeck{69.9} has to be considered. The major contribution to $S$
nevertheless originates from the doping by itself, this term being very
large due to the very small value of $x$. As the resistivity remains rather
large at high $T$, the $ZT$ values only increase up to 0.04 for $x =
0.03$. The thermal conductivity is another drawback for large $ZT$, with
$\kappa$ between 4 and \wkm{8} at 1000~K.  
Even if CuCrO$_2$ doped delafossites exhibit modest values of $ZT$, larger $ZT$
values have been obtained in CuRh$_{0.9}$Mg$_{0.1}$O$_2$ and
CuFe$_{0.99}$Ni$_{0.01}$O$_2$ \cite{Kur06,Noz07}, with a maximum of 0.15 at
1000~K and 0.14 at 1100~K respectively. In CuRh$_{0.9}$Mg$_{0.1}$O$_2$, a
smaller thermal conductivity ( $\simeq$ \wkm{1} at high T), leads to this
substantial enhancement of $ZT$ with respect to doped CuCrO$_2$ \cite{Kur06}.  

Large differences in the $ZT$ values are thus observed between CuCrO$_2$ and
these doped CuRhO$_2$ and CuFeO$_2$, and the role of Mg$^{2+}$ on doping in
CuCrO$_2$ was still unclear. Understanding these differences
\cite{Oku05,Ono07} has motivated the reinvestigation of transport properties
in CuCrO$_2$ and  CuRhO$_2$ doped with Mg$^{2+}$
\cite{Mai09a,Gui11,Mai09b}. 
X-ray diffraction combined with EDX analysis with transmission electron
microscopy has confirmed that the solubility of Mg$^{2+}$ is in fact very
restricted 
($x \simeq 0.10$ for CuRhO$_2$ and $x \simeq 0.03$ for CuCrO$_2$). In CuCrO$_2$,
the substitution by Mg$^{2+}$ rapidly leads to the formation of CuO (observed
from $x = 0.04$), and to the formation of the spinel MgCr$_2$O$_4$ as soon as
$x = 0.01$ \cite{Mai09b}. The evolution of the Seebeck coefficient as a
function of $x $ shows that doping is induced with the Mg$^{2+}$ substitution
even above $x = 0.01$ (formation of the spinel MgCr$_2$O$_4$) but is
suppressed for $x > 0.04$ as $S$ becomes constant. The transport properties
have thus been investigated up to 0.04. 

The magnetic structure is not strongly affected by the Mg$^{2+}$ substitution,
as revealed by neutron diffraction \cite{Poi09}. From magnetic
susceptibility, the high-spin state for Cr$^{3+}$ is confirmed and large
Curie-Weiss temperatures (-170~K) compared to a $T_N$ of 24~K (weakly affected
by Mg 
substitution) demonstrate the strong magnetic frustration associated to their
incommensurate antiferromagnetic structure. In the entire doping range,
electrical resistivity exhibits a localized behavior with $\frac{d\rho}{dT} <
0$, and large values of $\rho$ close to $10^{-1}$ – \Resist{10^2} at 300~K
depending on doping. Mg$^{2+}$ substitution leads to a reduction of $\rho$,
see the inset of Fig.~\ref{fig:Sylvie},
and for larger Mg$^{2+}$ content, $\rho$ can be measured down to low T,
with the magnetic transition directly observed at 24~K in the $\rho(T)$ curves. 
The Seebeck coefficient is positive, and evolves from a
localized behavior for $x = 0$ ($S \propto 1/T$, with very large values at room
temperature close to \seebeck{650}, see Fig.~\ref{fig:Sylvie}) to
smaller values for $x > 0$, with $\frac{dS}{dT} > 0$.
 
In the antiferromagnetic state, the resistivity and thermopower depend strongly 
on the magnetic field, as shown by the magnetothermopower and
magnetoresistance curves measured up to 9~T. As discussed earlier, the
introduction of Mg$^{2+}$ could both generate a mixed valency of
Cr$^{3+}$/Cr$^{4+}$ or Cu$^+$/Cu$^{2+}$. The Heikes formula which was used to
interpret the $S(x)$ dependence can not discriminate between these two
different doping origins. However, this existence of magnetoresistance and
magnetothermopower supports the Cr$^{3+}$/ Cr$^{4+}$ doping, with the
transport being dominated by the Cr – O network rather than the Cu network.  

At higher T, for $x > 0$, $S$ continuously increases up to 1100~K, while
$\rho$ continuously decreases, leading to power factor values close to 
\pfac{2.10^{-4}} \cite{Gui09}. This is very close to the values
previously reported \cite{Ono07}.  More discussion about the $S(T)$ curves
in connection to the 
band structure can be found in the Section~\ref{dft:curho2}. 
   
\begin{figure}[t!]
	\centering
	\includegraphics[width=\columnwidth]{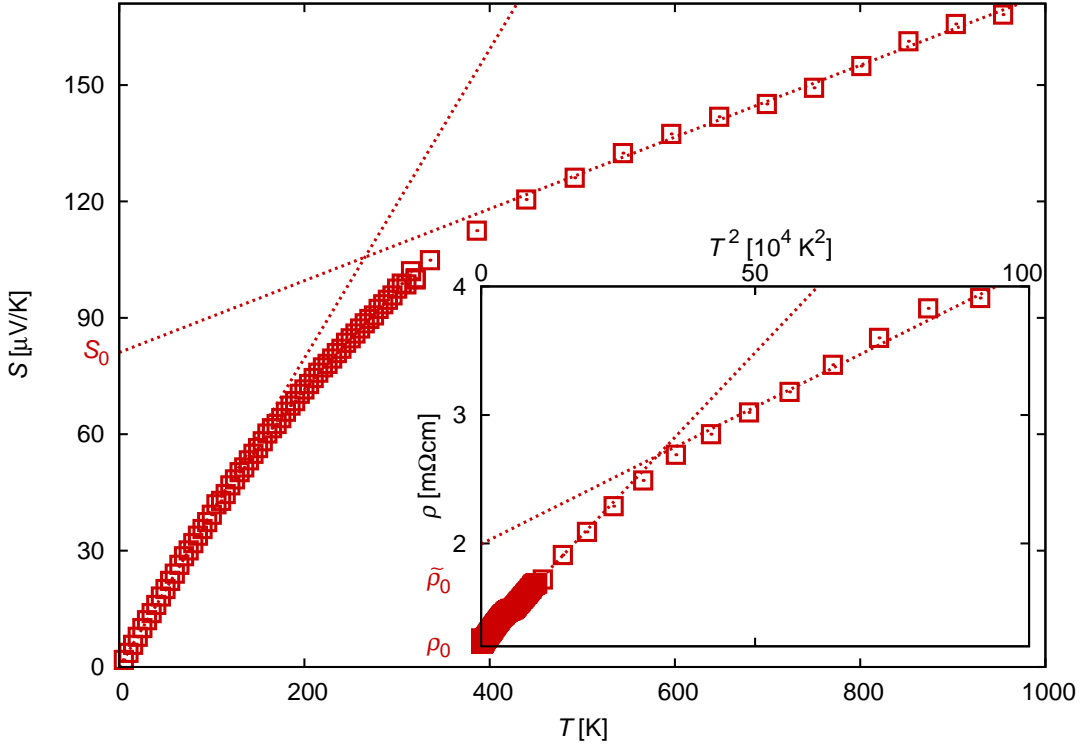}
	\caption{
Thermopower $S$ and resistivity $\rho$ (inset) of
CuRh${}_{0.9}$Mg${}_{0.1}$O${}_2$ as functions of the temperature $T$
(squares). The linear, respectively quadratic, regions are highlighted by the
dotted lines in order to show the transition from a behavior, which can be
explained within a Fermi liquid picture, towards a behavior which is referred
to as the one of an apparent Fermi liquid (AFL) and characterized by the
additional offsets $S_0$ and ${\tilde \rho}_0$. 
	}
	\label{fig:AFL}
\end{figure}

In the case of CuRhO$_2$, Mg$^{2+}$ substitution can reach
10\%, a value much larger than the one observed in the case of
CuCrO$_2$. Cu$_2$MgO$_3$ appears as an impurity for $x > 0.10$. From band
structure calculations, the Cu valency has been assigned to Cu$^+$, while Rh
is Rh$^{3+}$ in the low spin state.  
With Mg$^{2+}$ doping, the semiconducting behavior of CuRhO$_2$ is gradually
replaced by a metallic behavior, for $T > 100$~K in all samples. A minimum of 
resistivity is observed at $T_{\rm min}$, even for $x = 0.10$, with
$T_{\rm min}$ decreasing as $x$ increases. For large $x$, the
resistivity values reach the 
{\ensuremath{\textnormal{m}\Omega\,\textnormal{cm}}} values, typical of
the so-called ‘bad metallic’ 
oxides. Simultaneously, as shown in Fig.~\ref{fig:Sylvie}, the thermopower
decreases from large values 
(\seebeck{325} for $x= 0$), to values close to \seebeck{170} for $x =
0.10$. It must be emphasized that the Seebeck values for undoped CuCrO$_2$ are much
larger than the ones of CuRhO$_2$. For the latter, the spin entropy term
contributing to only $\sim$ \seebeck{70}, the larger $S$ difference can be
ascribed to a higher self-doping in CuRhO$_2$ naturally resulting from a small 
off-stoichiometry. 
The interesting
point is that the power factor $\frac{S^2}{\rho}$ presents a peculiar $T$
dependence, with an almost constant value from 300 to 1000~K for $x= 0.10$,
reaching \pfac{6. 10^{-4}}, a value typical of the best thermoelectric
oxides \cite{Mai09b}. This peculiar behavior comes from the $T^2$ behavior
observed for the 
electrical resistivity in a large $T$ range, associated to an almost $T$
linear behavior for $S$. The $S \propto T$ and $\rho \propto T^2$ behaviors
(see Fig.~\ref{fig:AFL}) 
have stimulated the development of the Apparent Fermi Liquid model
\cite{Kre12}.

Another proposed origin of the enhanced Seebeck values in Mg-doped CuRhO$_2$ is an electronic band structure consisting of a large flat region and a sharply dispersing edge \cite{Usu09}, a so-called 'pudding-mold' model. When the Fermi level is close to the transition between these two regions, the asymmetry can cause an enhanced Seebeck coefficient within the approximations of semiclassical transport.

All the above examples point out the rich variety of properties exhibited by
this family of 2D materials. In this ${\rm ABO_2}$ delafossite class of
compounds the pronounced two-dimensionality goes along with the triangular
arrangement of transition-metal ions with rather well localized electrons,
which has laid ground for the known variety of extraordinary phenomena. In
particular, when the B cation is a paramagnetic 3d metal, though there exists
geometric frustration in the antiferromagnetic exchange interactions, the
magnetic coupling through the separating A layer is sufficient to allow a
setting of a 3D antiferromagnetic ordering as in \cucro{} (or \pdcro{}). The
competition between the different in- and out-of-plane magnetic interactions is
responsible for very different antiferromagnetic structures, collinear in
\cufeo{} or non collinear in \cucro{}. In that respect, it is very difficult to
predict their spin driven multiferroic properties, ferroelectricity being
spontaneous at $T_N$ for \cucro{} but requiring the application of a magnetic
field to be induced in \cufeo{} \cite{Ter14}. For these antiferromagnetic
insulating delafossites with B$^{3+}$ cations, doping is necessary to induce
more electronic conducting states as the Mg$^{2+}$ for Cr$^{3+}$ substitution in
\cucro{}, with little effect on $T_N$. This behavior contrasts to that of
\curho{} with its fully occupied Rh $4d$ $t_{2g}$ ($S=0$) subshell which
neither displays strong localization effects nor magnetic order 
but serves as a possible thermoelectric material as it combines low electrical
resistivity and large Seebeck values. Finally, when the A cation is Pd or Pt
as in \pdcoo{} and \ptcoo{} (see Section~\ref{sec:Ramzy}), very different
electrical properties are observed, the samples behaving as “2D metals”
characterized by extreme anisotropy creating a unique situation among all
oxides. We point out that these five compounds serve only as paradigmatic
examples of the whole class of delafossites and the present survey is far from
being exhaustive. Indeed, the richness of phenomena can be further increased
by alloying and doping.

\section{Metallic delafossites}\label{sec:Ramzy}
\subsection{Introduction}
The majority of delafossite materials so far synthesized are semiconductors, 
of particular interest for their potential use as p-type transparent oxides. 
There is, however, a small group of materials that behave quite differently.

When the B site is occupied by Co, Rh or Cr, the \botwo{} layers have an
electronic configuration that makes them charged but formally insulating. The
octahedral crystal field around the B-site generates the familiar $t_{2g}$ and
$e_g$ bands, and the $t_{2g}$ bands are either completely filled by low-spin
$d^6$ Co$^{3+}$ or Rh$^{3+}$, or effectively so by magnetic high-spin $d^3$
Cr$^{3+}$. The resulting net single negative charge per formula unit requires
that the A-site ion carry a single positive charge.

If A$=$Ag$^+$ or Cu$^+$, the resulting electronic structure is typically
gapped, despite the fact that the Ag-Ag or Cu-Cu distance can be shorter than
in bulk Ag or Cu. On the other hand, the $d^9$ configurations of Pd$^+$ and
Pt$^+$ produce materials with extremely high, strongly anisotropic
conductivity. Conduction in the triangular A-plane is mediated by close
overlap between adjacent A-ions, which are separated by distances of
$2.830$~\AA, $2.923$~\AA, and $3.021$~\AA    $ $ 
for $\ensuremath{\textrm{Pd(Co,Cr,Rh)O}_2}$\cite{Sha71}. This compares to a nearest neighbor distance of $2.751$~\AA  $ $ in fcc palladium. Likewise in \ptcoo{}, the interatomic distance is $2.823$~\AA  $ $ \cite{Kus15}, compared to $2.775$~\AA $ $  in the metal. For comparison, the room temperature in-plane resistivities of \pdcoo{} and \ptcoo{} are $\resist{2.6}$\cite{Hic12} and $\resist{2.1}$\cite{Kus15}, much lower than for the fcc metals, which both have around \resist{10.6}. This dramatic reduction in resistivity for weaker overlap is counterintuitive.

Transport out-of-plane remains metallic in character, even though the separation between consecutive A-planes is as much as $6$~\AA. Thus \pdcoo{}, \ptcoo{}, \pdrho{}\cite{Kus17} and \pdcro{} are very good metals. \pdcro{} is additionally notable because of the frustrated antiferromagnetic ordering at 37~K. 

De Haas-van Alphen \cite{Hic12} as well as photoemission \cite{Noh09a} experiments show that the low-temperature Fermi surface of \pdcoo{} consists of a single electron-like warped hexagonal cylinder, while that of \pdcro{} resolves into electron- and hole-like bands that arise from magnetic reconstruction of the single sheet \cite{Ok13,Hic15}.

An exception to this scheme is \agnio{}, which also retains metallic
character \cite{Shi93}. In this case the $d^7$ Ni$^{3+}$ sets the Fermi energy in the $e_g$
band. As the temperature is reduced, orbital degeneracy is lifted by charge transfer between Ni sites and an ordered state arises when 3Ni$^{3+}$ $\rightarrow$ Ni$^{2+}$ + 2Ni$^{3.5+}$ and metallic conduction is preserved. This contrasts with other Ni$^{3+}$ materials where Jahn Teller distortions open a gap between the $e_g$ levels and lead to insulating ground states.

Metallic conductivity can also be obtained in many cases by doping, but these
are the only stoichiometric delafossites known to produce a metallic
ground state comparable to elemental metals. By contrast, the “bad metal” nature of the thermoelectric
delafossites discussed in the previous section requires electronic correlations to be taken into account. Below we focus on the transport properties of the clean
Pd/Pt materials, where the strong structural anisotropy and simple electronic
structure make them textbook cases of quasi two-dimensional metals and give
rise to effects rarely seen in the solid state. 

\subsection{Transport properties}
\subsubsection{Resistivity}
$ $\\[1em]
\pdcoo{}, \pdcro{} and \ptcoo{} are notable for their very low room
temperature in-plane resistivities, $\rho(\mathrm{300K}) =
\resist{2.6}$\cite{Hic12}, $\resist{10}$\cite{Tak10b} and $\resist{2.1}$
respectively \cite{Kus15}. This applies to transport confined to the ab-plane. Transport in the inter-plane direction maintains a metallic temperature dependence, although it is typically between 300 to 1000 times more resistive in nature \cite{Tak07,Hic12}. This indicates that there is still some degree of overlap between Pd/Pt planes, perhaps mediated by the \botwo{} layers.

\begin{figure}[t!]
\includegraphics[width=0.9\columnwidth]{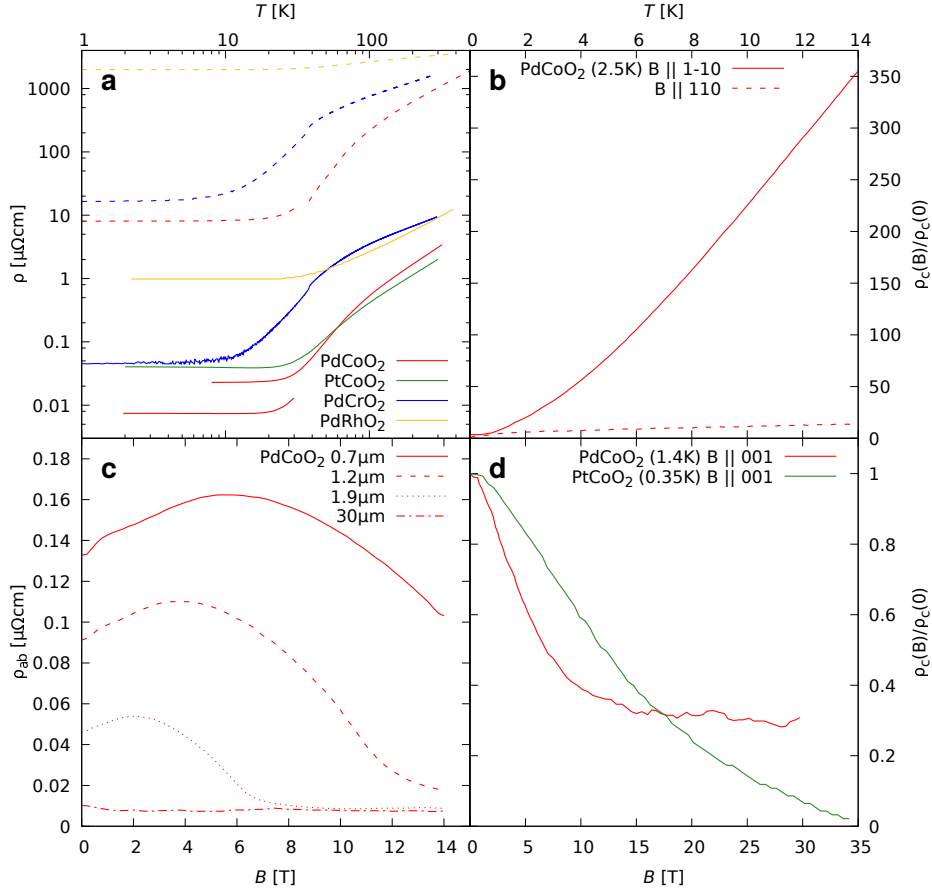}
\caption{Resistivity of metallic delafossites (compiled from
  Refs.~\cite{Tak07,Hic12,Dao15a,Kus15,Kus17,Mol16,Kik16}). a) The solid
  lines show the in-plane resistivity $\rho_{ab}$, while the dashed lines are
  out-of-plane resistivity $\rho_c$. $\rho_{ab}$ for two samples of \pdcoo{}
  of different residual resistivity are shown. At around 10~K both \pdcoo{}
  and \ptcoo{} have a shallow minimum in $\rho_{ab}$, and generally very
  similar temperature dependence and magnitude. The N\'eel transition in
  \pdcro{} is seen as a small kink in both transport directions. b)
  Transverse magnetoresistance $\rho_c(B)$ of \pdcoo{}. When the magnetic
  field is aligned along the $[1\bar{1}0]$ direction, there is a huge
  response. c) Channel width dependence of
  transverse magnetoresistivity $\rho_{ab}(B)$ in \pdcoo{} where the channel
  size is smaller than the mean free path. There is a complex dependence on
  magnetic field. d) Negative longitudinal magnetoresistance $\rho_c(B)$
  for \pdcoo{} and \ptcoo{}.}
\label{fig:rhoexp}
\end{figure}

The resistivity of \pdcoo{} has a super-linear temperature dependence not
characteristic of the usual electron-phonon scattering mechanisms (Fig.~\ref{fig:rhoexp}a). The
presence of optical phonon modes with a characteristic temperature of 250K was
suggested to account for this \cite{Tak07}. The form of the resistivity curve is similar in \ptcoo{} \cite{Kus15}.

Meanwhile in the magnetic analogue \pdcro{}, the resistivity has a sub-linear temperature dependence, and is around three times higher than in \pdcoo{} \cite{Tak10b}. This sub-linearity has been related to an extended range of magnetic scattering present related to the antiferromagnetic transition at $T_N = 37$~K.

The low temperature resistivity is of particular interest. In both \pdcoo{} and \ptcoo{} it falls to very low values, of the order of \resist{0.01}. A slight upturn occurs below 20K, with a magnitude of about 5\% of $\rho_0$ in \ptcoo{}. This feature is so far unexplained. There is no sign of magnetic impurities that might contribute to a Kondo effect in these materials. 

In \pdcoo{} the resistivity below 30~K can be fit to an exponentially
activated form (notwithstanding the very small upturn) \cite{Hic12}. This is argued to be a consequence of the simple electronic structure. Only one band crosses the Fermi energy, and the resulting Fermi surface is a closed cylinder that does not touch the Brillouin zone boundary. There is therefore a minimum phonon wavevector required for an electron-phonon umklapp scattering event to occur. Such phonons have a characteristic temperature of 30K, meaning that their population rises exponentially with this characteristic temperature. This leads to the observed exponentially activated resistivity. In the alkali metals, which also feature a single band Fermi surface entirely contained within the first Brillouin zone, the equivalent temperature is only  $\sim$4~K.

As a consequence of this suppression of electron-phonon umklapp scattering
combined with low intrinsic disorder, the mean free path can be of the order
of micrometers at low temperature. The result is the observation of
channel-size dependent resistivity \cite{Mol16} (Fig.~\ref{fig:rhoexp}c).
Since umklapp scattering is one of the few mechanisms that generates resistivity in clean systems, its suppression means that the electron fluid at low temperatures has few ways to dissipate the momentum acquired under an applied electric field. When boundary scattering becomes the dominant source of relaxation, the result is the viscous flow of electrons. It has even been suggested that if conditions are right, it may be possible to observe phenomena such as second sound in these metals at low temperatures.

The out-of-plane resistivity remains metallic (continuously increasing with temperature) for \pdcoo{} \cite{Tak07}, although the exponential activation at low temperatures is not present in this direction. This is as expected, as the Fermi surface touches the zone boundary in this direction and there are no forbidden electron-phonon interactions. The super-linear behavior at high temperature is also less pronounced than for the in-plane direction. 

In \pdcro{}, the onset of antiferromagnetic order causes a small cusp in the resistivity as the Fermi surface reconstructs. Below $T_N$, the resistivity approximately follows $T^3$. This rapid power law arises from the suppression of scattering in the ordered state. Sub-linear temperature dependence is evident for both in- and -out-of plane resistivity \cite{Tak10b}. Extrapolation of the magnetic susceptibility points to a Curie-Weiss temperature $\Theta_p\sim$500~K, an order higher than the N\'eel temperature. This implies a high degree of frustration. Specific heat \cite{Tak09} and neutron scattering \cite{Bil15} results point to the presence of magnetic fluctuations at temperatures between $T_N$ and $\Theta_p$. Magnon-electron scattering has been proposed to explain the elevated resistivity of \pdcro{} compared to non-magnetic \pdcoo{}.

\subsubsection{Magnetoresistance}
$ $\\[1em]
Strong transverse magnetoresistance (MR) is a characteristic of these
materials (Fig.~\ref{fig:rhoexp}b), and becomes more pronounced at low
temperatures. For in-plane current and magnetic field parallel to the c-axis,
the orbital motion of quasiparticles around the cylindrical Fermi surface
leads to MR of several hundred percent at low temperature. In absolute terms,
however, this magnetoresistance is only of the order of a few \resist{}. 

Striking results are obtained when the magnetic field is rotated in the plane
and out-of-plane transport is recorded. In this case MR can be several
thousand percent and depends sensitively on the in-plane field direction
\cite{Tak13}. Six-fold patterns are observed that can be reproduced well using
a semiclassical model of transport. Kohlers rule is also obeyed, implying that
a single relaxation mechanism is dominant. 

In the ordered state of \pdcro{} the localized moments undergo a spin-flop for
moderate magnetic fields applied in the plane \cite{Hic15}. A step is visible
in the magnetoresistance curve at around 6.5~T. The same kind of spin-flop
occurs at 5.3~T in semiconducting \cucro{}, where the antiferromagnetic
configuration is very similar. This emphasizes the local moment nature of the
magnetism; while coupling to the conduction electrons in \pdcro{} is
significant enough to open a gap, the magnetic ordering of the Cr$^{3+}$ is
largely unaffected by the presence of the conduction electrons.

Negative longitudinal magnetoresistance (LMR) has been observed in \pdcoo{} and \ptcoo{} (Fig.~\ref{fig:rhoexp}d) at low temperatures and for particular field angles \cite{Kik16}. The occurrence of negative LMR in several materials has been taken as evidence that they belong to a class of topological materials, the Weyl semimetals. The negative LMR in those cases arises from the counterflow of surface currents in the presence of parallel electric and magnetic fields. A similar but more generic effect has been proposed that occurs in clean materials when a warped cylindrical Fermi surface undergoes Landau quantization in strong magnetic fields \cite{Gos15}. Points with quasi-linear dispersion appear where the Landau tubes intersect the Fermi energy, and these play the role of the Weyl points, driving the counterflow of surface currents.

A less exotic origin of negative magnetoresistance in clean materials is the suppression of boundary scattering by strong fields when the sample dimension approaches the electronic mean free path. Cyclotron orbits smaller than the sample size can increase the distance electrons travel between collisions with the boundary, thus causing a reduction in resistance. This condition is satisfied in \pdcoo{} at low temperatures for typical samples, but it is not clear if it can account for the angular dependence of the observed magnetoresistance.

\subsubsection{Hall effect}
$ $\\[1em]
The Hall resistivity in \pdcoo{} and \ptcoo{} is non-linear with magnetic
field at intermediate temperatures, signaling the crossover from low- to
high-field behavior \cite{Tak10c, Kus15}. Values of $\omega_c\tau$ can greatly
exceed unity in magnetic fields accessible in the laboratory when aligned
parallel to the c-axis. The temperature dependence of the Hall coefficient is
generally monotonic, reflecting the single-band electronic structure, and the
values reached in the high-field limit match well with the expectation of one
electron per Pd/Pt ion. 

In \pdcro{}, with higher residual resistivity, $\omega_c\tau$ reaches
$O(1)$. A more complex temperature and magnetic field profile surrounds the
N\'eel transition. An early report interpreted this complexity as a sign of
the unconventional anomalous Hall effect \cite{Tak10c}, as there was no
obvious connection between the antiferromagnetic order parameter and magnitude
of the Hall coefficient. A subsequent study showed that a two-band model
incorporating magnetic breakdown could reasonably account for the
non-linearity of the Hall resistivity in the N\'eel state \cite{Ok13}. While
the proposed breakdown fields are quite small, the antiferromagnetic gap
opened in the Fermi surface is also found to be of the correct order of
magnitude. A study of the temperature dependence of the Hall coefficient close
to the N\'eel temperature revealed the impact of the short-range magnetic
fluctuations on macroscopic transport \cite{Dao15b}. Rather than being
dominated by changes in diffuse scattering by magnetic excitations, the Hall
coefficient smoothly interpolates between one-band and two-band electronic
structures as a result of coherent scattering between hotspots by magnetic
excitations. Hall effect data are shown in Fig.~\ref{fig:halltep}a. 

\subsubsection{Thermoelectricity}
$ $\\[1em]

\pdcoo{} and \ptcoo{} were suggested as interesting candidates for thermoelectric cooling on the basis of numerical calculations \cite{Ong10a}. These showed that the c-axis Seebeck coefficient could be as large as \seebeck{-200}, unusually large for a good metal. Combined with a limited electrical conductivity, the thermoelectric power factor $S^2/\rho$ could be large enough for applications to become feasible. 

\begin{figure}[t!]
\centering
\includegraphics[width=0.8\columnwidth]{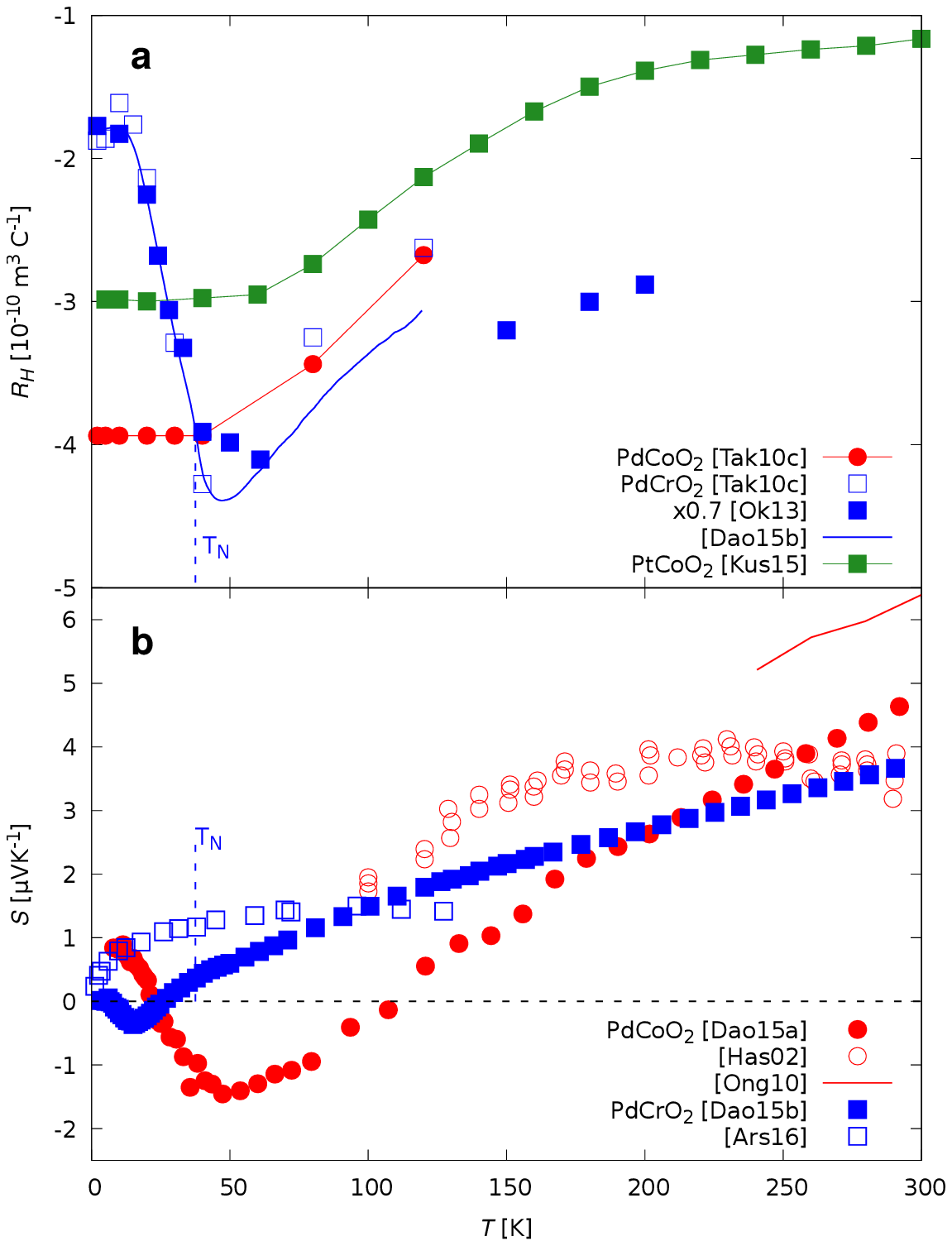}
\caption{a) Hall coefficient $R_H = \rho_{xy}/B$ of metallic delafossites in 8~T. Single band \pdcoo{} \cite{Tak10c} and \ptcoo{} \cite{Kus15} have a temperature dependent Hall coefficient as the mean free path increases to beyond the cyclotron radius when the temperature is decreased, causing a crossover from weak-field to strong-field behavior. In contrast, $R_H$ in \pdcro{}  \cite{Tak10c,Ok13,Dao15b} has a complex temperature dependence due to the compensation between bands at low temperature as well as the presence of magnetic fluctuations around $T_N$. b) Thermoelectric power of metallic delafossites in zero magnetic field. The thermoelectric power of \pdcoo{} single crystals \cite{Dao15a} shows a phonon drag peak at low temperature and a linear dependence with temperatures at high temperatures, as expected from calculations \cite{Ong10a}. Older polycrystalline samples \cite{Has02} may have been further from stoichiometry. There are also discrepancies between results on \pdcro{} single crystals \cite{Dao15b,Ars16}.}
\label{fig:halltep}
\end{figure}

In the planar direction, the predicted sign is positive and the magnitude is much smaller, which was in accord with an early report on polycrystalline \pdcoo{} \cite{Has02}. Owing to the tendency for single crystal samples to grow as thin platelets, it has so far not been possible to measure the out-of-plane Seebeck coefficient, however the measured in-plane values are in good agreement with the calculations \cite{Dao15a}. The strong predicted anisotropy and sign change of the Seebeck coefficient with respect to crystalline direction are another unique aspect of these materials. In one suggestion, this sign difference could be exploited in a monolithic type of thermoelectric device \cite{Zho13}. Zero-field data for \pdcoo{} and \pdcro{} are shown in Fig.~\ref{fig:halltep}b.

The temperature dependence of the in-plane Seebeck coefficient of \pdcoo{} is dominated by a linear component, as expected for a good metal. The slope of this component is close to the prediction of free-electron theory \cite{Ong10a}, around \sovert{0.018}. At low temperature there is a pronounced negative peak. 

This negative peak must be ascribed to phonon drag, whereby the flow of phonons along the thermal gradient exerts an additional pressure on the electrons. This effect is most pronounced at some fraction of the Debye temperature. The presence of phonon drag was confirmed by a study of the impact of sample purity on the size of the peak: cleaner samples yield a larger drag peak \cite{Dao15a}.

In the case of \pdcro{} the experimental situation is not yet clear; two recent studies produce rather different results for the temperature dependence of the Seebeck coefficient. Ref.~\cite{Dao15b} shows a Seebeck coefficient dominated by a linear term with a weak negative peak at low temperature, much like \pdcoo{}. The slope of the linear term is similar to that obtained in \pdcoo{}, as expected because of the similarities in electronic structure. The low temperature peak may arise from phonon drag or from a multi-band scenario as was used to explain the Hall effect data. In magnetic fields up to 8~T, the negative peak becomes more pronounced, which argues in favor of the multi-band scenario.

However, Ref.~\cite{Ars16} reports a Seebeck coefficient with a rather different temperature dependence. It also shows a strong suppression under intense magnetic fields, which is not seen in the moderate applied fields used in Ref.~\cite{Dao15b}. The non-linear form of the zero-field temperature dependence is somewhat unexpected, given the linear behavior in \pdcoo{}, whose electronic structure should be very similar at high temperatures. The suppression of magnon drag with increasing magnetic field was invoked to explain the strong magnetic field dependence. It is difficult to reconcile the greatly different data reported in these two studies; both crystals appear to be of similar quality and exhibit the same N\'eel transition.

\subsubsection{Thermal conductivity}
$ $\\[1em]

The anisotropy of the thermal conductivity of \pdcoo{} was extracted from a Montgomery-type experiment \cite{Dao15a}. The results ruled out any possible application in energy conversion; the thermoelectric efficiency would be negligible because of a significant out-of-plane lattice thermal conductivity of around \wkm{50} at room temperature. The in-plane thermal conductivity is dominated by the electronic contribution and reaches $\sim$\wkm{300} at 300~K, but with a Lorenz ratio approaching 1.1. This suggests that the lattice contribution to the thermal conductivity is much more significant than in the noble metals, for example, where the Lorenz ratio does not usually rise above unity.

\subsubsection{Nernst effect}
$ $\\[1em]
The Nernst effect was studied across the antiferromagnetic transition in
\pdcro{} \cite{Dao15b}. The Nernst effect is the thermoelectric equivalent of
the Hall effect and is sensitive to changes in mobility and carrier number. In
\pdcro{}, the Nernst coefficient also responds to the presence of the
antiferromagnetic fluctuations that foreshadow long-range order. The response
can be understood within the same framework as the Hall effect, where the
opening of the gaps at the hotspots is smoothed by the presence of
fluctuations. In the ordered, two-band state the Nernst coefficient is large
and depends strongly on magnetic field, while in the paramagnetic single-band
phase it is small and of a magnitude compatible with a high carrier
density. The crossover region is characterized by a sign change and a strong
increase in magnitude. 

\subsubsection{Discussion}
$ $\\[1em]
The simplicity of the electronic structure in \pdcoo{} and \ptcoo{} leads to
interesting transport properties even in the absence of strong
correlations. These materials are self-organising on the nanoscale,
alternating conducting and insulating layers. In the case of \pdcro{}, the
insulating layer also supports local moment magnetism. This unique material
permits the study of symmetry breaking on transport properties without
contaminating the conduction layer with magnetic ions.  

\section{Theoretical analysis}\label{sec:theo_ana}
\subsection{Electronic structure}
\label{dft}

Since its invention about five decades ago, density functional theory 
has witnessed a tremendous success story in predicting, explaining, and 
understanding the electronic properties of matter. This overwhelming  
progress was initiated by the fundamental work of Hohenberg, Kohn, 
and Sham, who established the electronic density as the key variable 
to access the properties of the ground state \cite{Hoh64,Koh65}. 
Nowadays, density functional theory is an integral and indispensable part 
of condensed matter research and materials science, which impressively 
complements experimental studies and has found its way into industrial 
laboratories \cite{Wim16,Roz16}. 

Since calculations as based on density functional theory, usually 
called first principles calculations, do not need any input data 
other than the atomic numbers of the constituent atoms and their 
initial coordinates, they serve as an ideal and independent starting 
point for any investigation of materials. While summarizing recent 
first principles work, we largely follow the discussion presented 
in previous publications by our group, which should be consulted for a 
detailed account of all results \cite{Eye08a,Eye08b,Mai09a,Mai09b}. 
In the present context we mention only that all calculations were 
performed using the augmented spherical wave (ASW) method, which likewise 
is described in some detail elsewhere \cite{Eye00,Eye13}. 
\subsection{$ {\rm PdCoO_2} $ and $ {\rm PtCoO_2} $}
\label{dft:pdcoo2}

A number of electronic structure investigations of these two 
Co-delafossites have been reported in the literature.  
Most of them focused on the extraordinary conductivity and specifically 
tried to clarify the composition of the wave functions at the Fermi 
energy. In particular, linear muffin-tin orbital calculations were 
performed by Seshadri {\em et al.}\ as well as by Okabe {\em et al.}\ 
\cite{Ses98,Oka03}. The former authors, who also investigated 
$ {\rm PtCoO_2} $, attributed the density of states at $ {\rm E_F} $ 
mainly to the Pd $ 4d $ states with only small contributions from the 
Co $ 3d $ and O $ 2p $ orbitals. In contrast, photoemission data 
were interpreted assuming the density of states at the Fermi energy arises
exclusively from the Pd $ 4d $ states \cite{Tan98,Hig98,Mar06}. 
Furthermore, from the combination of photoemission spectroscopy and 
inverse photoemission spectroscopy it was concluded that the Fermi energy 
is located in a shallow minimum of the density of states and doping may 
thus cause rather high values of the thermoelectric power 
\cite{Hig98,Has02,Hig04}. Hence, an investigation as that described below 
was needed to resolve the controversy by identifying the influence of the 
different atomic species and orbitals on the electronic properties and 
thereby to make closer connection with the photoemission data \cite{Eye08a}. 
In contrast to previous calculations in the literature, which were based 
on the crystal structure data by Shannon, Rogers, and Prewitt \cite{Sha71}, 
fully optimized lattice parameters and atomic positions as given in 
Table\ \ref{dft:pdcoo2:tab} were used. 
\begin{table}[b!]
\caption{Experimental and calculated lattice parameters (in \AA) and 
         atomic positions.}
\begin{tabular}{llccc}
compound          &        & a      &    c    & $ z_O $ \\
\hline
$ {\rm PdCoO_2} $ & exp.\  & 2.8300 & 17.743  & 0.1112 \\
                  & calc.\ & 2.8767 & 17.7019 & 0.1100 \\
$ {\rm PtCoO_2} $ & exp.\  & 2.8300 & 17.837  & 0.1140 \\
                  & calc.\ & 2.8989 & 17.458  & 0.1128 \\
\end{tabular}
\label{dft:pdcoo2:tab}
\end{table}

\begin{figure}[t!]
\centering
\includegraphics[width=0.9\columnwidth,clip]{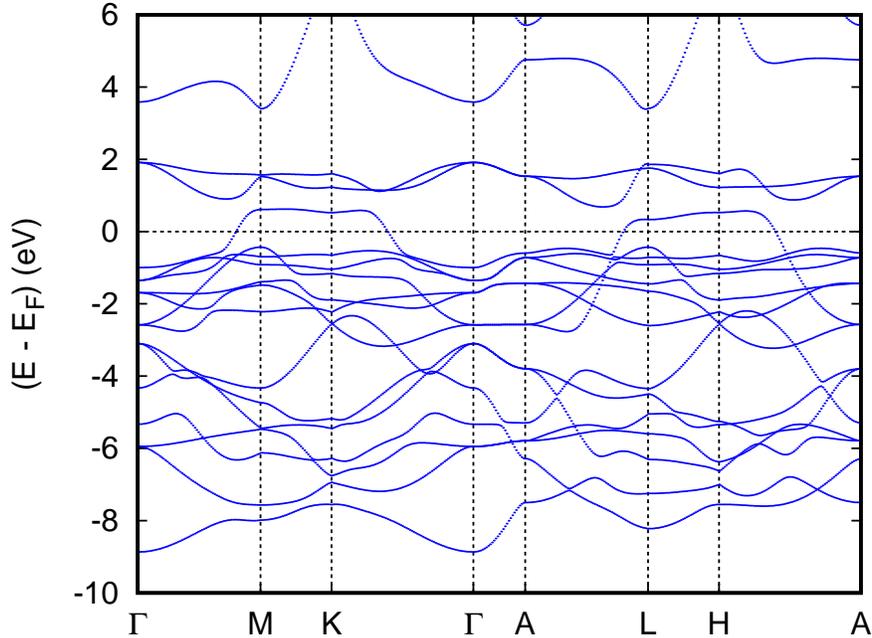}
\caption{Electronic bands of $ {\rm PdCoO_2} $. Reprinted with permission from
(Chem. Mater. \textbf{20}, 2370 (2008)). Copyright (2008) American Chemical
Society.}  
\label{dft:pdcoo2:bnd}
\end{figure}

\begin{figure}[t!]
\centering
\includegraphics[width=0.8\columnwidth,clip]{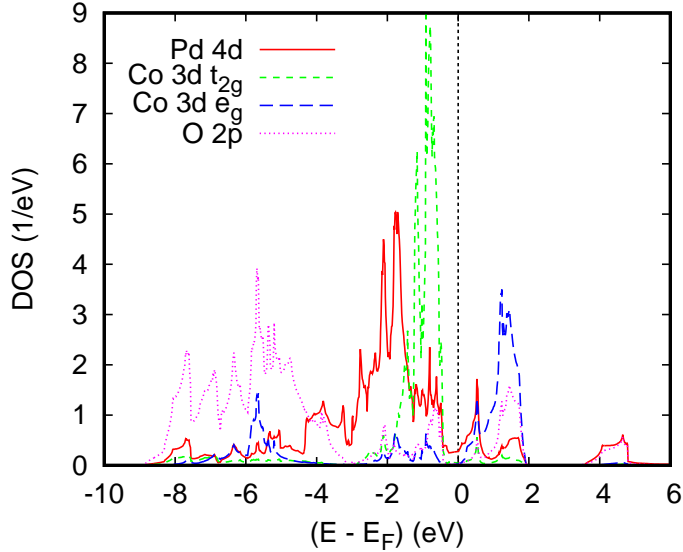}
\caption{Partial densities of states (DOS) of $ {\rm PdCoO_2} $.
         Selection of Co $ 3d $ orbitals is relative to the local 
         rotated reference frame, see text. Reprinted with permission from
(Chem. Mater. \textbf{20}, 2370 (2008)). Copyright (2008) American Chemical
Society.}
\label{dft:pdcoo2:dos}
\end{figure}

In discussing the electronic properties of the two cobaltates we start 
displaying the electronic bands along high-symmetry lines of the first 
Brillouin zone of the hexagonal lattice as well as the partial densities 
of states (DOS) in Figs.\ \ref{dft:pdcoo2:bnd} 
and \ref{dft:pdcoo2:dos}, respectively.
The somewhat complicated structure of both the electronic bands and the 
DOS results from the energetical overlap of the relevant orbitals in 
the energy interval shown. Yet, close to the Fermi energy this 
complexity clears up completely and only a single band straddles 
$ {\rm E_F} $, leading to a very simple Fermi surface. Since a detailed 
discussion of the results can be found in Ref.\ \cite{Eye08a}, we here 
focus on the most important findings. These include the crossover from 
predominant O $ 2p $ bands to a set of sharp peaks due to transition 
metal $ d $ states at about $ -4 $ eV and a clear separation of the 
Co $ 3d $ states into their $ t_{2g} $ and $ e_g $ manifolds due to 
the octahedral coordination by oxygen atoms. Note that the latter 
observation refers to a local rotated coordinate system with the 
Cartesian axes pointing approximately towards the oxygen atoms. 
Since the Fermi energy falls right between the $ t_{2g} $ and $ e_g $ 
manifolds, Co is found in a $ d^6 $ low-spin state. In this respect 
$ {\rm PdCoO_2} $ is thus not unlike $ {\rm CuRhO_2} $ to be considered 
below. However, while in the latter compound the $ d $ orbitals are fully 
occupied and, hence, allow for the semiconducting behavior, the missing 
electron in $ {\rm PdCoO_2} $ leads to incomplete band filling and the 
finite conductivity. In conclusion, the above results confirm the 
picture of trivalent Co and monovalent Pd in a $ d^9 $ configuration 
\cite{Sha71,Tan98,Mar06}. At the same time, they clearly reveal the 
only tiny contribution of the Co and O states to the electrical 
conductivity, which is carried almost exclusively by the Pd $ 4d $ states. 

\begin{figure}[t!]
\centering
\includegraphics[width=0.8\columnwidth,clip]{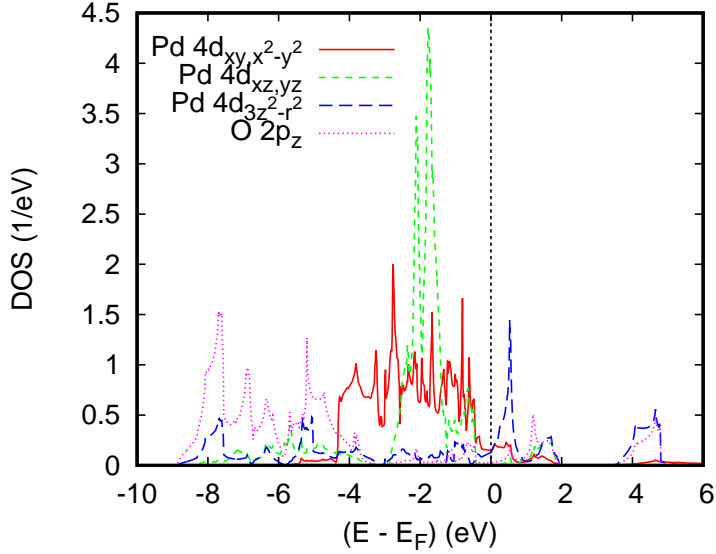}
\caption{Partial Pd $ 4d $ and O $ 2p_z $ DOS of $ {\rm PdCoO_2} $. Reprinted
  with permission from (Chem. Mater. \textbf{20}, 2370 (2008)). Copyright (2008) American Chemical
Society.} 
\label{dft:pdcoo2:dos-d}
\end{figure}

It is very instructive to further analyze the latter in terms of their 
five $ 4d $ partial DOS, which are shown in Fig.~\ref{dft:pdcoo2:dos-d}.
Since Pd is linearly coordinated by two oxygen atoms parallel to the 
$ c $ axis and has six Pd neighbors in the $ a $-$ b $ plane, to this 
end the global coordinate system is used. With this choice, contributions 
from the $ d_{xy} $ and $ d_{x^2-y^2} $ as well as from the $ d_{xz} $ 
and $ d_{yz} $ states are identical. The Pd $ 4d $ partial DOS are 
strongly influenced by the linear coordination. 
Again, without going 
into details, we mention the strong $ \sigma $-type 
$ d_{3z^2-r^2} $-$ p_z $ overlap along the $ c $ axis as indicated by 
the striking similarity of the corresponding partial DOS in the 
energy range from $ -9 $ to $ -4 $\,eV as well as the broad 
Pd $ d_{xy,x^2-y^2} $ bands reflecting the short in-plane Pd-Pd 
distances, which are very close to those of metallic Pd. Obviously, 
these latter states and the $ d_{xy,x^2-y^2} $ states add the largest 
contributions to the total DOS at $ {\rm E_F} $, whereas that of the 
$ d_{xz,yz} $ states is almost negligible. Finally, the sharp peak 
of the $ d_{3z^2-r^2} $ partial DOS at about $ + 0.6 $\,eV can be 
clearly assigned to the almost dispersionless band along the lines 
M-K and L-H. 

\begin{figure}[t!]
\centering
\includegraphics[width=0.8\columnwidth,clip]{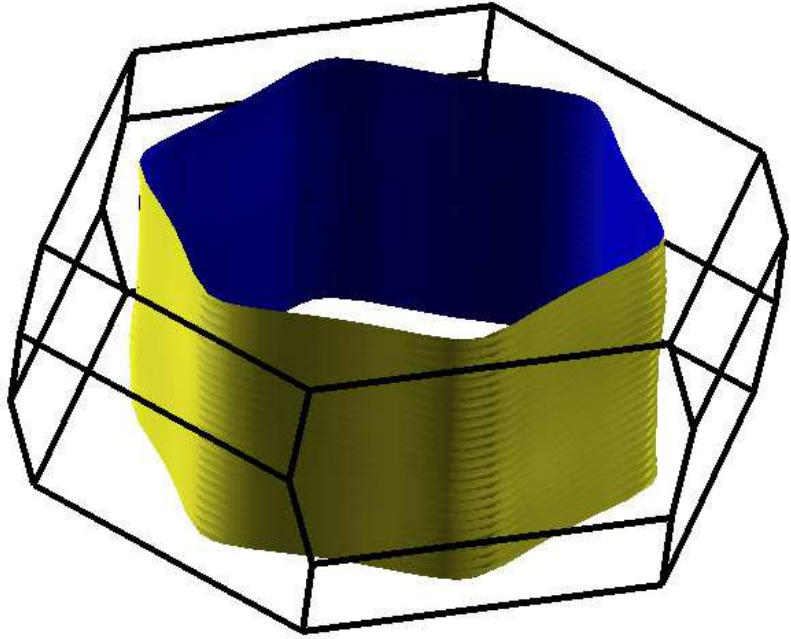}
\caption{Fermi surface of $ {\rm PdCoO_2} $. Reprinted with permission from
(Chem. Mater. \textbf{20}, 2370 (2008)). Copyright (2008) American Chemical
Society.}
\label{dft:pdcoo2:frm}
\end{figure}

Finally, the Fermi surface shown in Fig.~\ref{dft:pdcoo2:frm} 
underlines the strong quasi two-dimensionality of the electronic states 
and, hence, explains the strong anisotropy in electric conductivity. 

The physical picture emerging from the above results agrees very well 
both with previous calculations \cite{Ses98,Oka03} and more recent 
first-principles studies \cite{Ong10a,Ong10b,Gru15}. They are also in 
agreement with photoemission and x-ray absorption data by Tanaka 
{\em et al.}, Higuchi {\em et al.}, and Noh {\em et al.}\ 
\cite{Tan98,Hig98,Noh09a,Noh09b}, who likewise attribute the metallic 
conductivity almost exclusively to the Pd $ 4d $ states and even regard 
\pdcoo{} as a metal-insulator stack structure \cite{Noh09a}. 
These authors attribute the high conductivity to the strong dispersion 
of the conduction band, the large Fermi surface, and the long lifetime 
of the charge carriers. The extraordinary transport properties of \pdcoo{} 
were studied by Takatsu {\em et al.}, by Ong {\em et al.}, by Gruner 
{\em et al.},\ as well as by Daou {\em et al.}, who reported on the strong
anisotropy of the electrical 
conductivity \cite{Tak07,Dao15a} and the thermoelectric power \cite{Ong10a,Gru15}.
This goes along with de Haas-van Alphen measurements of Hicks {\em et al.}, 
who found anomalously low contributions from electron-phonon, 
electron-electron, and electron-impurity scattering to the resistivity 
\cite{Hic12}. Finally, the exceptional magnetoresistance has been discussed  
by Takatsu {\em et al.}\ as well as Kikugawa {\em et al.}\ \cite{Tak13,Kik16}. 

There is a dispute about the detailed composition of the 
electronic wave functions at the Fermi surface. Arguments in favor 
of hybridized Pd $ d_{3z^2-r^2} $-$ 5s $ orbitals were supported 
by Kimura {\em et al.}\ \cite{Kim09}. However, 
the above results demonstrate that the metallic conductivity is maintained 
by the in-plane $ d_{xy} $ and $ d_{x^2-y^2} $ orbitals and the in-plane 
part of the $ d_{3z^2-r^2} $ orbitals to a similar degree with a somewhat 
greater influence of the former. 

In passing, we mention very similar findings for $ {\rm PtCoO_2} $. 
Yet, the anisotropy of the electrical conductivity is reduced in this 
compound due to the larger extent of the Pt $ 5d $ orbitals and the 
resulting increased overlap of the in-plane $ d $ states as well as 
with the O $ 2p $ states.

\subsection{$ {\rm CuFeO_2} $}
\label{dft:cufeo2}

As already mentioned before, one of the most attractive properties of 
the delafossites arises from the arrangement of transition-metal 
ions on a two-dimensional triangular lattice, which allows to study, 
{\em e.g.}, frustration effects, incommensurate and commensurate 
non-collinear magnetic order, unusual spin coupling, and multiferroic 
behavior. $ {\rm CuFeO_2} $ and $ {\rm CuCrO_2} $ have found much 
interest in this context and are therefore also considered in this 
overview. 

The investigations of magnetic delafossite compounds has been from 
the very beginning accompanied by the search for distortions 
occurring at low temperatures in order to escape the geometric 
frustration. For this reason, the exact atomic and magnetic 
structure of many of these compounds has long been a matter of 
extensive dispute. This is not different for $ {\rm CuFeO_2} $ 
\cite{Mui67,Dou86}. Furthermore, there has been also some dispute 
about the possible occurrence of multiple magnetic phase transitions. 
From neutron diffraction data, Mekata {\em et al.}\ identified 
two magnetic transition for $ {\rm CuFeO_2} $ at 
$ {\rm T_{N1} = 16} $\,K and $ {\rm T_{N2} = 11} $\,K. The magnetic 
phases required monoclinic and orthorhombic magnetic supercells 
of the undistorted rhombohedral unit cell with commensurate and 
incommensurate collinear arrangements of the localized 
$ 4.4 \mu_B $ $ {\rm Fe^{3+}} $ moments \cite{Mek92,Mek93,Pet00,Kim06}. 
More importantly, using x-ray and neutron diffraction measurements, 
Ye {\em et al.}\ indeed found structural distortions below 
4\,K accompanying the magnetic phase transitions and leading to a 
monoclinic lattice with space group $ C2/m $ \cite{Ye06}. In addition, 
analysis of spin-wave spectra gave strong hints at a three-dimensional 
magnetic coupling \cite{Ye07}. It was also found that in response to 
a magnetic field the magnetic transition temperatures lower and an 
incommensurate structural distortion as well as ferroelectricity is 
induced \cite{Ye06}. Observation of a noncollinear-incommensurate phase 
in magnetic field was likewise taken as indicative of multiferroic 
behavior by Kimura {\em et al.}\ \cite{Kim06} and later on indeed identified 
in Al-doped $ {\rm CuFeO_2} $ \cite{Sek07}. Further indication of 
multiferroic behavior was taken from inelastic neutron scattering data 
\cite{Ye07}. Ruttapanun {\em et al.}\ have pointed to the 
potential of Pt-doped $ {\rm CuFeO_2} $ for thermoelectric applications 
\cite{Rut11}. 

Despite strong interest, only few electronic structure calculations 
for magnetic delafossite compounds had been reported 
\cite{Gal97,Ses98,Ong07,Sin07}. This is possibly due to the still 
much debated atomic and magnetic structure. For that reason, only 
the ferromagnetic configuration was considered. Yet, the results 
were contradictory. Galakhov {\em et al.}\ reported on a ferromagnetic 
ground state for the rhombohedral $ R\bar{3}m $ structure with a 
magnetic moment at the Fe site of about 0.9\, $ \mu_B $, much lower 
than the experimental value \cite{Gal97}. Furthermore, the Fe $ 3d $ 
$ t_{2g} $ states were found above the Cu $ 3d $ states just at the 
Fermi energy, again in disagreement with both photoemission data and 
the fact that $ {\rm CuFeO_2} $ is a semiconductor with an optical 
band gap of about 1.15\,eV. In contrast, LDA$+U$ calculations revealed 
a band gap of 2\,eV and a magnetic moment of 3.76\,$ \mu_B $. However, 
the occupied Fe $ 3d $ states were located at about 9\,eV below the 
valence band maximum and thus much too low \cite{Gal97}. More recently, 
Ong {\em et al.}\ found in their calculation a high-spin state with a magnetic
moment of  
3.78\,$ \mu_B $ per Fe and the Fe $ 3d $ $ t_{2g} $ spin-up states 
below the Cu $ 3d $ bands in agreement with photoemission and x-ray 
emission data \cite{Ong07}. However, again a finite optical band gap 
was arrived at only after taking into account local electronic 
correlations within the GGA$+U$ scheme. 

\begin{figure}[t!]
\centering
\includegraphics[width=0.8\columnwidth,clip]{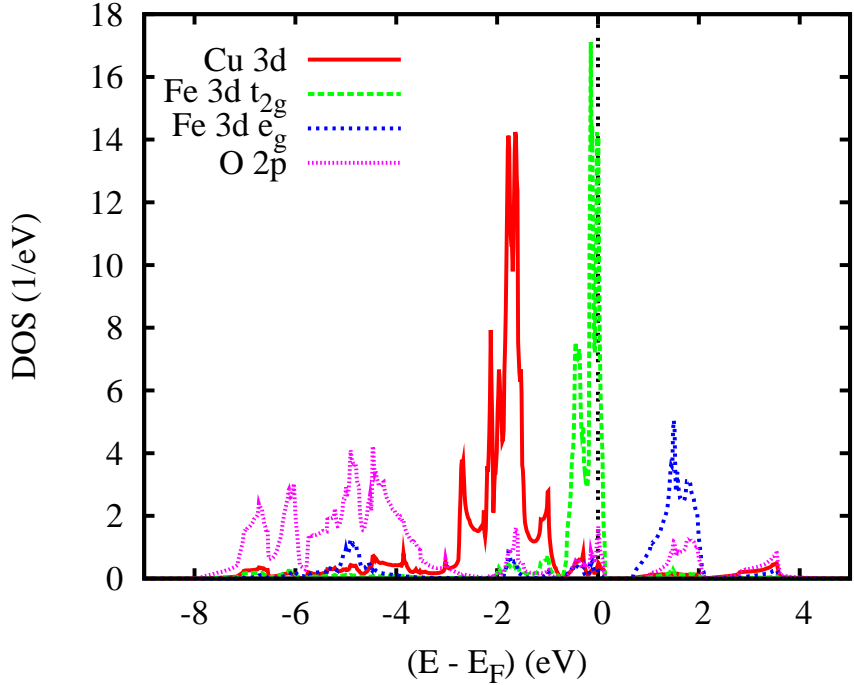}
\caption{Partial densities of states (DOS) of rhombohedral $ {\rm CuFeO_2} $.
  Selection of Fe $ 3d $ orbitals in this and the subsequent figures is
  relative to the local rotated reference frame, see text. Reprinted with
  permission from (Phys. Rev. B \textbf{78}, 052402 (2008)). Copyright (2008)
  American Physical Society.}
\label{dft:cufeo2:dosnm}
\end{figure}

We begin a detailed analysis by discussing the density of states (DOS) 
arising from spin-degenerate calculations for the rhombohedral structure 
as displayed in Fig.~\ref{dft:cufeo2:dosnm}.
Crystal structure data by Ye {\em et al.} \cite{Ye06} are used throughout. 
As for $ {\rm PdCoO_2} $, a crossover from dominating O $ 2p $ states 
to a sequence of sharp transition metal $ d $ states is observed at about 
$ -3 $\,eV. While Cu is found in a monovalent $ d^{10} $ configuration 
in good agreement with experiment, Fe assumes a $ d^5 $ state with the 
Fermi energy falling into the upper part of the $ t_{2g} $ manifold. 
Note that in distinguishing the $ t_{2g} $ and $ e_g $ manifolds we 
again used the local rotated coordinate system with the Cartesian 
axes adjusted to an assumed perfect oxygen octahedron. Details can 
be found in Ref.~\cite{Eye08b}. 

In order to be in line with the previous work by Galakhov {\em et al.}\ 
as well as by Ong {\em et al.}\ \cite{Gal97,Ong07}, we next mention 
spin-polarized calculations for an assumed ferromagnetic state. Three 
different configurations were obtained corresponding to a low-spin, 
intermediate-spin, and high-spin moment located at the Fe site. The 
total energies as compared to the spin-degenerate configuration 
and the local magnetic moments are summarized in Table\ \ref{dft:cufeo2:tab}.
\begin{table}[b!]
\caption{Total energies (in mRyd per formula unit) and magnetic 
         moments (in $ \mu_B $) for different crystal structures 
         and magnetic orderings of $ {\rm CuFeO_2} $.}
\begin{tabular}{ccrrr} 
structure  & magn.\ order & $ \Delta E $ & $ m_{\rm Fe} $ & $ m_{\rm O} $ \\
\hline
rhombohedral\    & spin-deg.\   & $      0.0 $ &                &               \\
rhombohedral\    & ferro (LS)   & $    -16.7 $ & $       1.03 $ & $    - 0.02 $ \\
rhombohedral\    & ferro (IS)   & $    -12.0 $ & $       2.02 $ & $    - 0.02 $ \\
rhombohedral\    & ferro (HS)   & $    -19.2 $ & $       3.73 $ & $      0.21 $ \\
monoclinic & spin-deg.\   & $     -6.0 $ &                &               \\
monoclinic & ferro (LS)   & $    -21.5 $ & $       1.04 $ & $    - 0.02 $ \\
monoclinic & ferro (IS)   & $    -19.0 $ & $       2.08 $ & $    - 0.02 $ \\
monoclinic & ferro (HS)   & $    -32.0 $ & $       3.62 $ & $      0.19 $ \\
monoclinic & antiferro    & $    -46.0 $ & $   \pm 3.72 $ & $  \pm 0.08 $ \\
\end{tabular}
\label{dft:cufeo2:tab}
\end{table}
All three ferromagnetic configurations are lower in energy than the 
spin-degenerate case with the high-spin state being the most stable 
as long as the lattice is restricted to be rhombohedral. The results 
thus confirm both the low-spin and high-spin calculations by Galakhov 
{\em et al.}\ as well as by Ong {\em et al.}. In particular, they 
assign the differences between their findings to the existence of 
difference spin states. 

In a second step, the monoclinic structure observed by Ye {\em et al.}\ 
is considered \cite{Ye06}. Since this monoclinic unit cell still 
contains one Fe atom it allows only for spin-degenerate and 
spin-polarized ferromagnetic calculations. The resulting total energies 
and magnetic moments are given in Table\ \ref{dft:cufeo2:tab}. While 
the latter are very similar to those obtained for the rhombohedral 
structure, the total energies are all lower by several mRyd with the 
largest energy lowering occurring for the high-spin state. In passing, 
we mention that the similarities between the results obtained for both 
structures extend also to the partial densities of states. 

\begin{figure}[t!]
\centering
\includegraphics[width=0.8\columnwidth,clip]{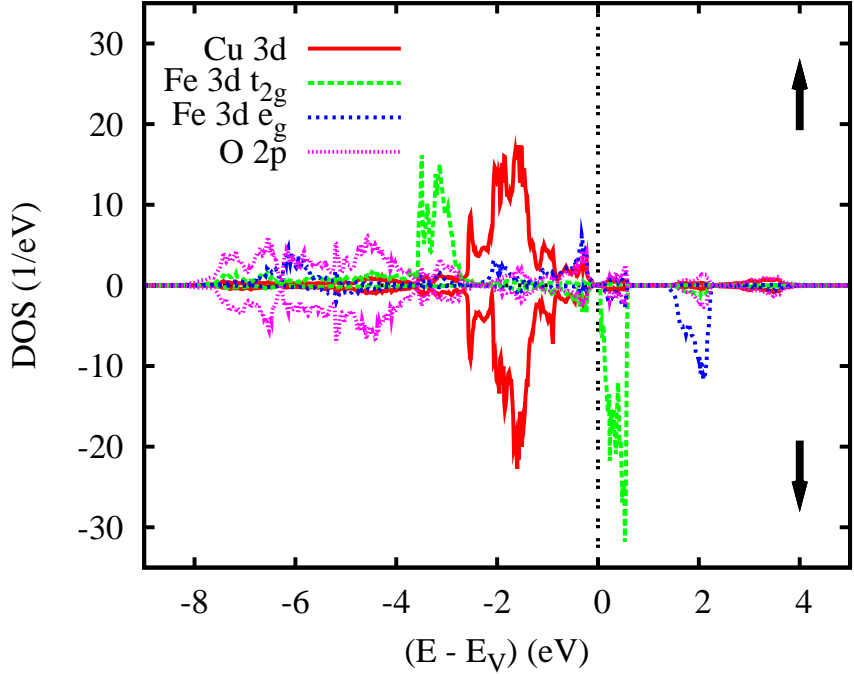}
\caption{Partial densities of states (DOS) of monoclinic antiferromagnetic
  high-spin $ {\rm CuFeO_2} $. Reprinted with
  permission from (Phys. Rev. B \textbf{78}, 052402 (2008)). Copyright (2008)
  American Physical Society.}
\label{dft:cufeo2:dosaf}
\end{figure}

Finally, we turn to calculations for the eightfold magnetic supercell 
proposed by Ye {\em et al.}\ \cite{Ye06}. The resulting partial DOS 
are displayed in Fig.~\ref{dft:cufeo2:dosaf} 
and the local magnetic moments and total energy included in Table\ 
\ref{dft:cufeo2:tab}. Obviously, the antiferromagnetic state 
has the lowest energy as compared to all other configurations and Fe 
is found to be in a high-spin state in agreement with the neutron 
diffraction data by Mekata {\em et al.}\ \cite{Mek92,Mek93}. Moreover, the
calculation yields a band gap of 0.05\,eV, in contrast to all previous results 
for $ {\rm CuFeO_2} $. The results for the ferromagnetic high-spin 
states as well as the antiferromagnetic ground state have been nicely 
confirmed by Zhong {\em et al.}, who, in addition, discussed the 
multiferroic behavior in terms of the hybridization of the Fe $ 3d $ 
and O $ 2p $ states \cite{Zho10}. In order to obtain a band gap in 
closer agreement with experiment, Hiraga {\em et al.}\ performed 
GGA$+U$ calculations and obtained a band gap of about 1\,eV for 
$ U = 4 $\,eV \cite{Hir11}.

\subsection{$ {\rm CuCrO_2} $}
\label{dft:cucro2}

$ {\rm CuCrO_2}$ is yet another example for strong geometric frustration 
effects coming with the predominantly antiferromagnetic coupling of rather 
well localized magnetic moments on a triangular lattice. However, unlike 
$ {\rm CuFeO_2}$, $ {\rm CuCrO_2}$ does not show any deformation of the 
atomic structure down to lowest temperatures. The magnetic structure has 
been at the center of controversial debates for a long time. While early 
neutron powder diffraction work was in favor of an out-of-plane 
$ 120^{\circ} $ spin structure with a commensurate propagation vector 
along the $ (\frac{1}{3},\frac{1}{3},0) $ direction \cite{Kad90}, 
more recent studies revealed two possible structures, namely, a helicoidal 
and a cycloidal structure with incommensurate propagation vector 
$ (0.329,0.329,0) $ below $ T_N = 24 $ K \cite{Poi09}. More recently, 
single-crystal polarized neutron diffraction found the spiral plane 
to be parallel to the $ (110) $ plane \cite{Sod09}. Building on x-ray 
diffraction measurements, Kimura {\em et al.}\ reported on a slight 
deformation of the triangular lattice plane accompanying the magnetic 
ordering \cite{Kim09b}. However, this finding has not been confirmed by other
groups. 

According to neutron powder diffraction data, magnetic susceptibility 
measurements, electrical permittivity and electrical polarization data 
of magnetically diluted $ {\rm CuCr_{1-x}M_xO_2} $ with 
$ {\rm M = Al, Ga, Sc, Rh} $ by Pachoud {\em et al.}\ the magnetic 
ground state turned out to be very robust against removal of Cr
\cite{Pac12}. These authors thus concluded that the magnetic ground 
state of $ {\rm CuCrO_2} $ is very different from that of $ {\rm CuFeO_2} $.  

In order to arrive at a better understanding of this material both 
spin-degenerate and spin-polarized calculations using the crystal 
structure data reported by Crottaz {\em et al.} \cite{Cro96} were performed. 
As for the other compounds discussed above, the partial densities of 
states as resulting from the spin-degenerate calculations have the 
lower part of the spectrum dominated by O $ 2p $ states, whereas the 
transition metal $ d $ states show rather sharp peaks above $ -4 $\,eV. 
While Cu again is found in a monovalent $ d^{10} $ configuration in 
close analogy with experimental findings, the Cr $ 3d $ states fall 
into half-occupied $ t_{2g} $ and empty $ e_g $ manifolds. From the 
$ d^3 $ configuration and the fact that $ E_F $ is very close to the 
highest peak one would expect long-range ferromagnetic ordering of 
Cr moments of 3 $\mu_B$ in a spin-polarized calculation. 

In view of the complex magnetic structure observed for $ {\rm CuCrO_2} $ 
it is useful to start spin-polarized calculations by first considering 
an assumed ferromagnetic order. This procedure is well motivated by the 
previous work on $ {\rm CuFeO_2} $, where we also started investigating 
an assumed ferromagnetic high-spin state before performing calculations 
for the antiferromagnetic ground state proposed by Ye {\em et al.}\ and 
found a strong similarity of the partial densities of states arising 
from these two configurations. Hence, we can learn a lot about the 
local electronic properties already from studying the ferromagnetic state. 

\begin{figure}[t!]
\centering
\includegraphics[width=0.8\columnwidth,clip]{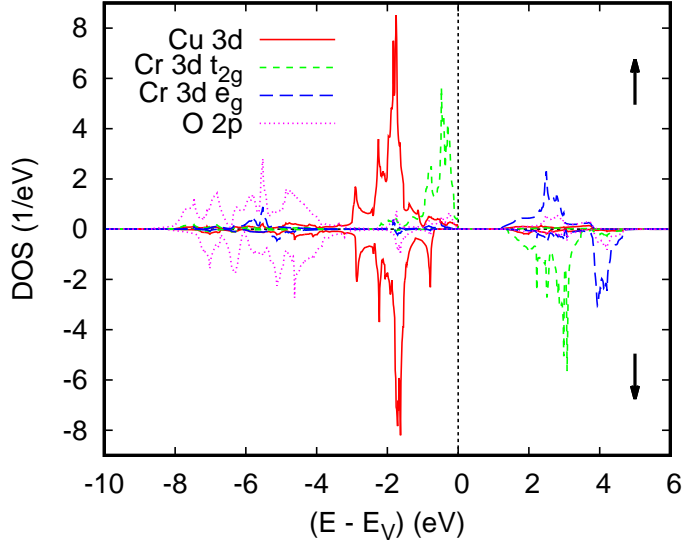}
\caption{Partial densities of states (DOS) of rhombohedral ferromagnetic 
         $ {\rm CuCrO_2} $. 
Reprinted from Solid State Commun. \textbf{149}, A.~Maignan, C.~Martin,
R.~Fr\'esard, V.~Eyert, E.~Guilmeau, S.~H\'ebert, M.~Poienar, and
D.~Pelloquin, \textit{On the strong impact of doping in the triangular
  antiferromagnet ${\rm CuCrO_2}$}, 962-967, Copyright (2009) with permission
from Elsevier.}
\label{dft:cucro2:dosfe}
\end{figure}

Now, from spin-polarized calculations for an assumed ferromagnetic 
configuration of $ {\rm CuCrO_2} $, a stable ferromagnetic configuration
was obtained with magnetic moments of 3.0\,$ \mu_B $. Most importantly,
the density of states as displayed in Fig.~\ref{dft:cucro2:dosfe} 
reveals a fundamental band gap of about 1.2 eV between the spin-up and
spin-down Cr $ 3d $ $ t_{2g} $ states, which also carry the overwhelming
part of the magnetic moment. Except for the lower filling of the 
Cr $ 3d $ $ t_{2g} $ states the partial densities of states look very 
similar to those of $ {\rm CuFeO_2} $, which fact saves us a detailed 
discussion here. 

\begin{figure}[t!]
\centering
\includegraphics[width=0.8\columnwidth,clip]{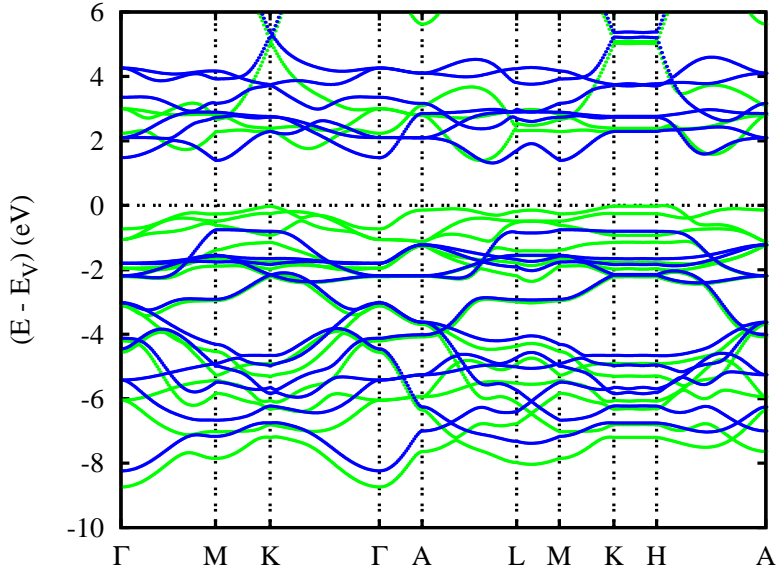}
\caption{Electronic bands of rhombohedral ferromagnetic $ {\rm CuCrO_2}$. 
Green (blue) curves correspond to the majority (minority) spin bands.
Reprinted from Solid State Commun. \textbf{149}, A.~Maignan, C.~Martin,
R.~Fr\'esard, V.~Eyert, E.~Guilmeau, S.~H\'ebert, M.~Poienar, and
D.~Pelloquin, \textit{On the strong impact of doping in the triangular
  antiferromagnet ${\rm CuCrO_2}$}, 962-967, Copyright (2009) with permission
from Elsevier.} 
\label{dft:cucro2:bndfe}
\end{figure}

The electronic bands along selected high-symmetry lines of the first 
Brillouin zone of the hexagonal lattice as displayed in 
Fig.~\ref{dft:cucro2:bndfe} 
reveal substantial three-dimensional dispersion, which we attribute 
to the coupling between the layers. Yet, the dispersion is considerably 
reduced close to the valence band maximum. In addition, the highest 
occupied states at M and K are almost identical to those at L and H. 
Hence, there is almost no dispersion of these bands along the lines 
M-L and K-H. As a consequence, within a rigid band approximation, we 
would expect strongly localized bands arising from small hole doping, 
which would induce finite but still small Fermi velocities driving 
spin-dependent transport. 

Performing GGA$+U$ calculations for a variety of delafossites,  
Scanlon and coworkers considered an antiferromagnetic 
structure for $ {\rm CuCrO_2} $ with ferromagnetic alignment 
within the sandwich planes and antiferromagnetic alignment of 
neighboring planes \cite{Sca09,Sca10,Sca11}. In addition, 
hybrid functional calculations were performed. While the authors 
also obtained a band gap separating occupied $ d $ states from 
empty Cr $ 3d $ $ e_g $ bands of about 1, 2, and 3 eV from their 
GGA, GGA$+U$, and hybrid functional calculations, respectively, 
their results are at some variance with those presented above. 
In particular, Cu $ 3d $ and Cr $ 3d $ $ t_{2g} $ states are both 
found in the same energy interval ranging from about $ -3 $ eV to 
the valence band maximum and bands close to the latter mainly trace 
back to the Cu states \cite{Sca09,Sca10,Sca11}. A very similar order 
of bands was obtained by Hiraga {\em et al.}\ \cite{Hir11}. 

\begin{figure}[t!]
\centering
\includegraphics[width=0.8\columnwidth,clip]{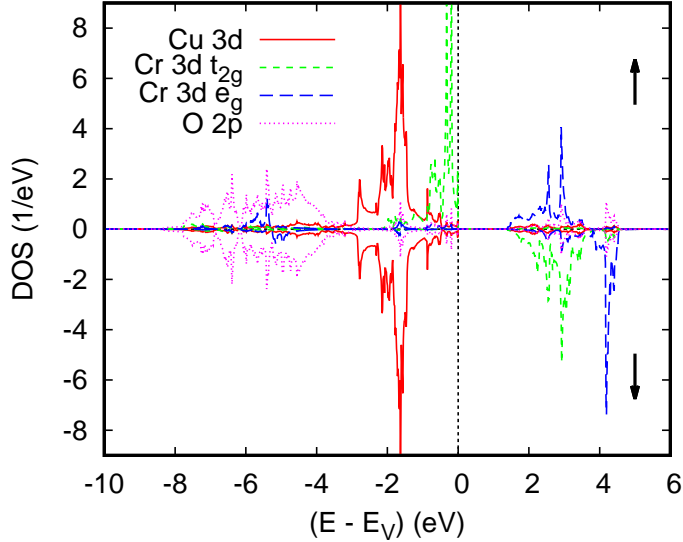}
\caption{Partial densities of states (DOS) of hexagonal antiferromagnetic 
         $ {\rm CuCrO_2} $.}
\label{dft:cucro2:dosaf}
\end{figure}

\begin{figure}[htb]
\centering
\includegraphics[width=0.8\columnwidth,clip]{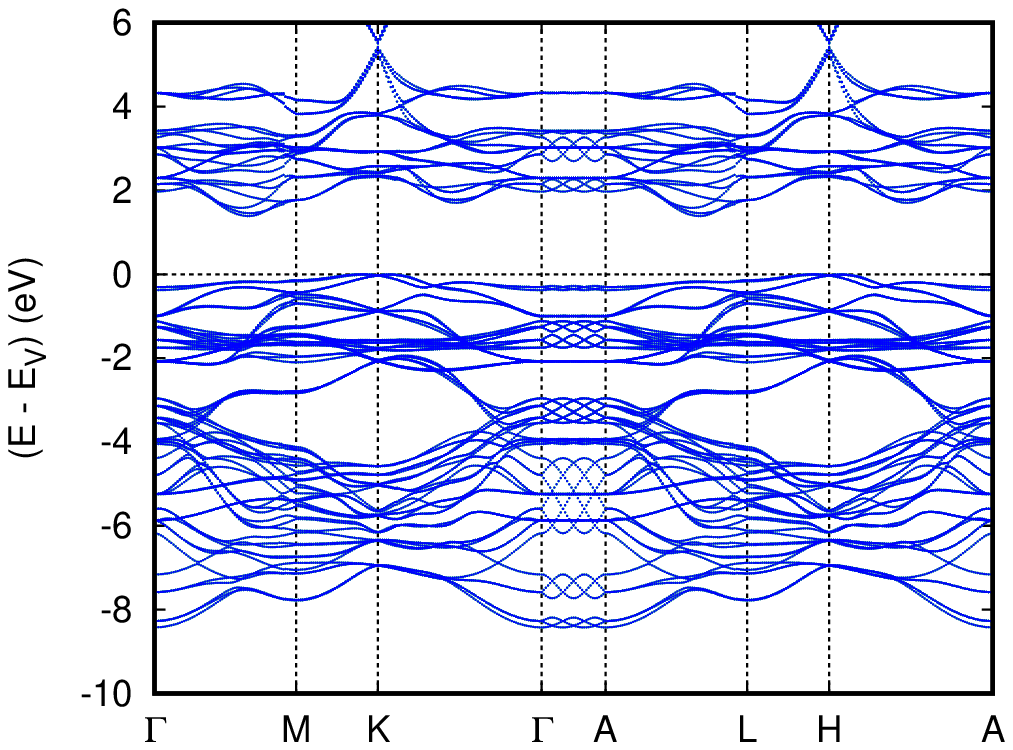}
\caption{Electronic bands of hexagonal antiferromagnetic $ {\rm CuCrO_2}$.}  
\label{dft:cucro2:bndaf}
\end{figure}

In order to arrive at a more complete picture, we complemented the 
above calculations for the assumed ferromagnetic state with 
calculations for the antiferromagnetic structure proposed by 
Scanlon and coworkers. This structure requires a hexagonal rather 
than rhombohedral unit cell with a doubling of the hexagonal $ c $ 
axis and it comprises six formula units. The resulting partial 
densities of states and electronic bands are display in 
Figs.~\ref{dft:cucro2:dosaf} 
and Fig.~\ref{dft:cucro2:bndaf}. 
Note that while representing the electronic band structure we used the 
Brillouin zone of the original small cell and, hence, obtain threefold 
band oscillations along the line $ \Gamma $-A. Apart from that, both 
the band structure and the partial densities of states are very similar 
to those obtained for the assumed ferromagnetic order and thus confirmed 
our above expectations, namely, that the electronic properties are 
already very well described by those for an assumed ferromagnetic state. 
Still, especially the spin-majority Cr $ 3d $ $ t_{2g} $ bands at 
the valence band maximum are much more localized in the antiferromagnetic 
configuration. Nevertheless, the discrepancy as compared to the results 
by Scanlon {\em et al.}\ are not yet resolved. As a consequence, the 
predominant character of the wave function occurring on small hole doping 
is not known. 

The issue has been critically discussed from both an experimental 
and a theoretical point of view by Yokobori {\em et al.}, who combined 
photoemission spectroscopy, soft x-ray absorption spectroscopy, and 
electronic structure calculation within the LDA$+U$ approximation as applied 
to the rhombohedral ferromagnetic configuration \cite{Yok13}. These authors 
confirmed the above results with the Cr $ 3d $ $ t_{2g} $ dominating the 
valence band maximum and the fully occupied Cu $ 3d $ bands found well 
below. Nevertheless, their x-ray absorption spectra on 
$ {\rm CuCr_{1-x}Mg_xO_2} $ showed a strong sensitivity to the Mg content 
indicating the hole will be doped into the Cu sites in contradiction 
to their photoemission spectra. In order to resolve the issue, 
Yokobori {\em et al.}\ proposed strong Cu $ 4s $-Cr $ 3d $ charge transfer 
via the O $ 2p $ states. Nevertheless, this puzzling situation is still 
awaiting further insight. 

Calculations of Jiang {\em et al.}, which took into account non-collinear 
spin arrangements, revealed predominantly in-plane exchange interactions 
and an incommensurate spin-spiral structure with a (110) spiral plane and 
a screw-rotation angle close to $ 120^{\circ} $ in agreement with 
experimental data \cite{Jia12,Sod09}. While spin-orbit interaction was 
shown to have only minor influence on the electronic properties spin 
frustration had a stronger effect on the $ d $-$ p $ hybridization. 
Finally, Monte Carlo simulations using exchange parameters extracted 
from supercell calculations led to a N\'{e}el temperature of 29.9\,K, 
again in acceptable agreement with the experimental findings
\cite{Jia12}. Recent Monte Carlo simulations taking small
lattice distortions into account yield a N\'{e}el temperature of $T_N \simeq
27\,K$. Furthermore, a connection between the emergence of spin helicity below
$T_N$ and ferroelectricity could be established \cite{Alb17}. 

\subsection{$ {\rm CuRhO_2} $}
\label{dft:curho2}

Finally, we turn to yet another delafossite compound, which has recently 
attracted much interest from a completely different perspective, namely, 
as a very promising candidate for applications in thermoelectricity and 
water splitting \cite{Kur06,Shi06,Mai09b,Usu09,Gu14,Toy15}. Kuriyama 
{\em et al.}\ and Shibasaki {\em et al.}\ observed a room-temperature 
thermopower of 130\,$ {\rm \mu V K^{-1}} $ and of 70\,$ {\rm \mu V K^{-1}} $, 
respectively, for $ {\rm CuRh_{0.9}Mg_{0.1}O_2} $ \cite{Kur06,Shi06}. 
In addition, the former authors reported a figure of merit 
$ ZT \approx 0.15 $ at 1000\,K \cite{Kur06}. 

There were reports on the use of $ {\rm CuRhO_2} $ 
as a photocathode for water splitting under visible light 
\cite{Gu14,Toy15}. Application in this field benefits from the fact 
that the band edges are optimally adapted to the water oxidation and 
reduction redox potentials \cite{Gu14}. Gu {\em et al.}\ attributed 
the stability of the delafossites as photocathodes to the fact that 
the electron acceptor levels are strongly dominated by the B-type 
atoms, {\em i.e.}, the Rh $ d $ levels and pointed to the high 
sensitivity of the stability and efficiency to even small Cu $ 3d $ 
contributions in this energy range \cite{Gu14}. Thus, there was and 
still is thus a very high motivation to understand the electronic 
properties of $ {\rm CuRhO_2} $ in detail. 

Our calculations were based on the crystal structure data by Oswald 
{\em et al.}\ \cite{Osw89}, who determined the lattice constants 
as $ a = 3.08 $\,{\AA} and $ c = 17.09 $\,\AA. Since these authors 
did not measure the internal oxygen parameter, we performed a 
structural optimization leading to a value of $ z_{\rm O} = 0.10717 $, 
which was used in all subsequent calculations
\cite{Mai09b}. 

\begin{figure}[t!]
\centering
\includegraphics[width=0.8\columnwidth,clip]{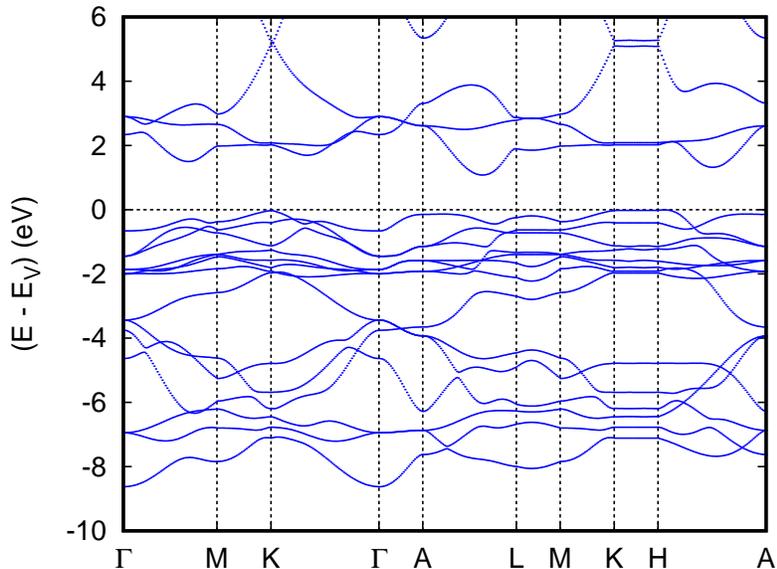}
\caption{Electronic bands of $ {\rm CuRhO_2} $. Reprinted with
  permission from (Phys. Rev. B \textbf{80}, 115103 (2009)). Copyright (2009)
  American Physical Society.}
\label{dft:curho2:bnd}
\end{figure}

\begin{figure}[tb]
\centering
\includegraphics[width=0.8\columnwidth,clip]{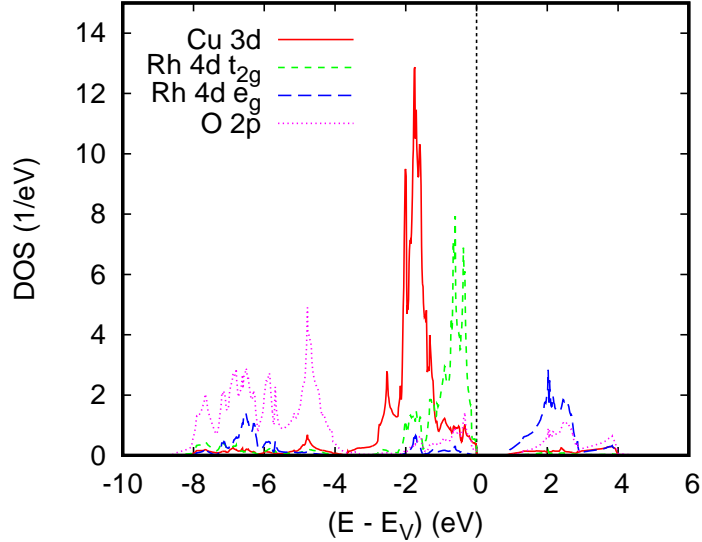}
\caption{Partial densities of states (DOS) of $ {\rm CuRhO_2} $. Selection of
  the Rh $ 4d $ orbitals is relative to the local rotated reference frame, see
  text. Reprinted with permission from (Phys. Rev. B \textbf{80}, 115103
  (2009)). Copyright (2009) American Physical Society.}
\label{dft:curho2:dos}
\end{figure}

The electronic bands along selected high-symmetry lines of the first
Brillouin zone of the hexagonal lattice and the partial densities of 
states (DOS) are displayed in Figs.~\ref{dft:curho2:bnd} 
and \ref{dft:curho2:dos}, respectively. 
Again, we observe the well known order of states with the crossover 
from O $ 2p $-dominated bands to sharp transition-metal peaks at 
about $ -4 $\,eV. As for the other Cu-based delafossites discussed 
above, Cu is found in a monovalent $ d^{10} $ configuration in 
close analogy with the experimental findings with the major $ 3d $ 
peaks well below the valence band edge. The Rh $ 4d $ states clearly 
exhibit splitting into a fully occupied $ t_{2g} $ manifold and 
empty $ e_g $ bands, which are separated by the fundamental band 
gap. The calculated value of the latter is about 0.85\,eV, which is 
somewhat smaller than the experimental value of 1.9\,eV. Rh is thus 
found in a $ d^6 $ configuration. The situations is not unlike that 
of $ {\rm CuCrO_2} $, where, however, only the spin majority $ t_{2g} $ 
states are occupied and the spin minority $ t_{2g} $ bands are shifted 
upwards to join the $ e_g $ bands above the band gap. 

It is interesting to note the band finite dispersion parallel to 
$ \Gamma $-A indicative of considerable three-dimensionality arising 
from the coupling between the layers. This is contrasted with the 
barely noticeable dispersion particularly along the line K-H. Yet, 
both the finite dispersion along $ \Gamma $-A and the flat bands 
along K-H have been also observed for the other delafossite materials.  

\begin{figure}[htb]
\centering
\includegraphics[width=0.8\columnwidth,clip]{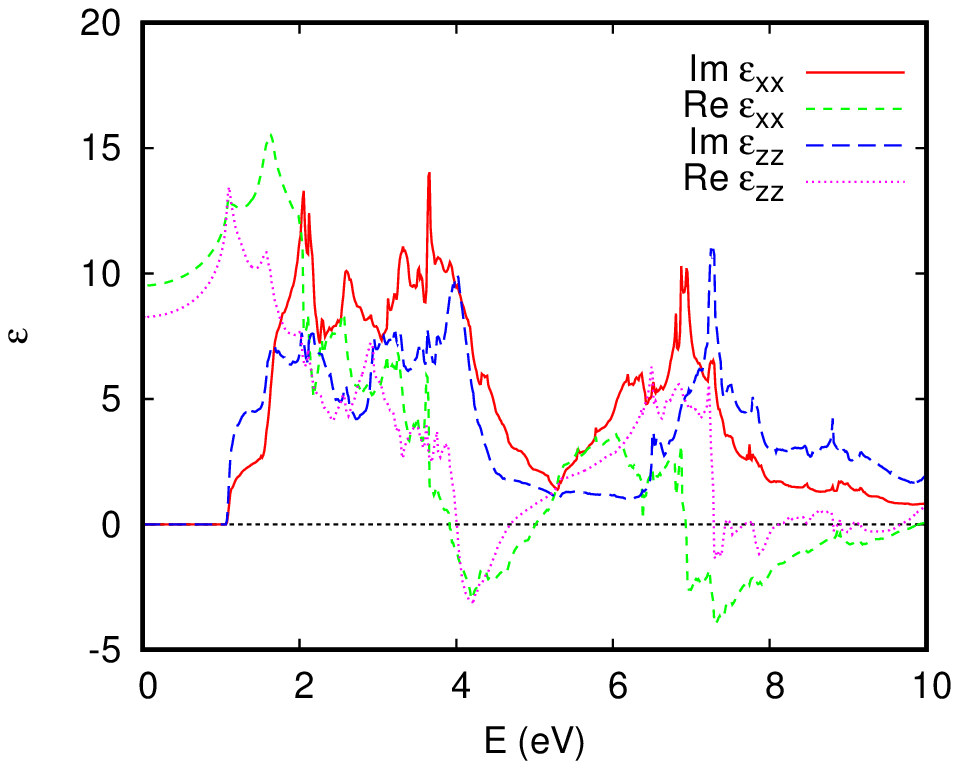}
\caption{Dielectric function of $ {\rm CuRhO_2} $.}
\label{dft:curho2:eps}
\end{figure}

Further insight about the electronic properties is obtained from the 
optical spectra. The real and imaginary parts of the dielectric 
function as calculated within linear-response (see Ref.\ \cite{Eye13} 
for more details) are shown in Fig.~\ref{dft:curho2:eps}. 
Obviously, the asymmetry between the in-plane and out-of-plane 
directions is not reflected in the absorption gap following from 
the imaginary part of the dielectric function. In fact the gap 
is very close to 0.85\,eV in all three directions, which is most 
likely to exceed the Hund's rule coupling. Therefore, the low-spin 
$ 4d^6 $ configuration of $ {\rm Rh^{3+}} $ is expected to be the 
ground state. This is consistent with earlier findings by Singh 
for $ {\rm CuCoO_2} $, where  the Co ions adopt the low-spin 
$ 3d^6 $ configuration \cite{Sin07}. 

\begin{figure}[htb]
\centering
\includegraphics[width=0.8\columnwidth,clip]{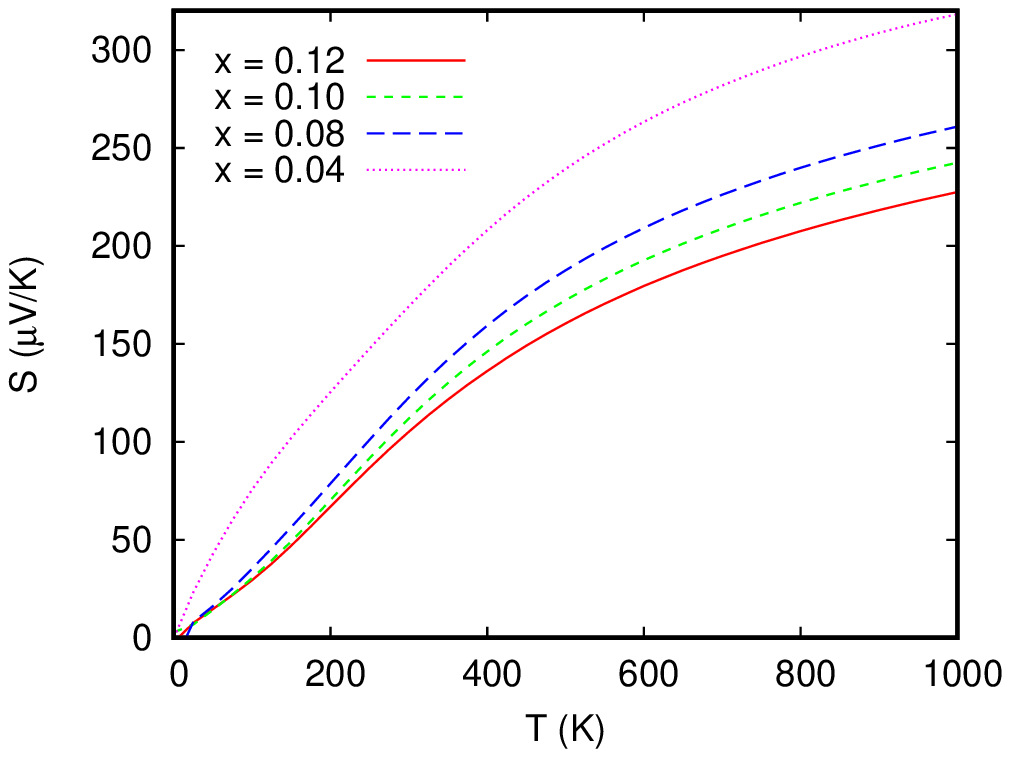}
\caption{Thermopower $ S $  of $ {\rm CuRhO_2} $ for different hole doping 
         levels. The curves display the average of all three diagonal 
         components of the tensor.} 
\label{dft:curho2:tep}
\end{figure}

Finally, we discuss the thermoelectric properties as evaluated from 
Boltzmann theory \cite{All96,Eye13}. The transport properties are 
usually expressed in terms of the Onsager transport coefficients 
(see also Ref.~\cite{Mai09b} for details). The calculated Seebeck 
coefficients for different hole doping levels are displayed in  
Fig.~\ref{dft:curho2:tep}. 
Note that the curves show the average of all components of the diagonal 
of the respective tensor. As expected, the thermopower strongly decreases 
with increased hole doping. In addition, it shows the characteristic 
bending towards smaller increase at high temperatures. As concerns 
especially the shape of these curves, the results are in almost perfect 
agreement with the experimental data as shown in Fig.~\ref{fig:AFL} of our own
group, the calculations of Usui {\em et al.}\  
\cite{Usu09}, who likewise used the Boltzmann approach, and with the 
experimental data by Kuriyama {\em et al.}\
\cite{Kur06}. Nevertheless, 
the absolute scaling of the thermopower is at variance both between the 
calculations and between different experiments. This should be related 
to the fact that calculated Seebeck coefficients are rather sensitive to 
details of the crystal structure. In addition, in applying Boltzmann 
theory, all calculations performed so far used a rigid band picture in 
order to mimic doping effects. Yet, a doping level of 0.1 holes per 
formula unit is not a small perturbation. For this reason, 
$ {\rm CuRh_{0.9}Mg_{0.1}O_2} $ should be regarded a metal rather than 
a doped semiconductor as is indeed done in the literature. Nevertheless, in view of these 
limitations, the above results are remarkably convincing.

\section{Conclusions and perspectives} \label{sec:concl}

It is remarkable that going from $d^9$ to $d^{10}$ monovalent cations at the A
site of A$^+$M$^{3+}$O$_2$ delafossite the electronic groundstate could switch
from “ultra” metallicity as in \pdcoo{} to semiconductors for \cucro{} or
\curho{}. Accordingly, very different physics in this class of materials can
be studied as illustrated by a few examples leading for instance to
multiferroicity when A is a $d^{10}$ monovalent cation in the case of \cufeo{}
and \cucro{}. Their triangular lattices of localized magnetic moments at the M
site is responsible for magnetic frustration which is lifted by
antiferromagnetic ordering at rather low N\'eel temperatures of $T_N = 16~K$
and $T_N = 24~K$ for \cufeo{} and \cucro{}, respectively. For both compounds,
the electronic spin-polarized structure calculations are consistent with the
experimental observations that the antiferromagnetic semiconductors states are
the most stable with high-spin states for Fe$^{3+}$ or Cr$^{3+}$. These
calculations also reveal energy gaps at $E_F$ fitting with the poor conducting
behavior. However, the limit of these calculations lie in the prediction of
the magnetic structure for which subtle changes in the in-plane and
out-of-plane exchange energies lead to very different antiferromagnetic
structures below $T_N$ as the 4SL collinear structure in \cufeo{} and the
incommensurate structures, helicoidal or cycloidal in \cucro{}. 
Another challenge for the calculations is created by the prediction of the thermoelectric properties. In these $d^{10}$ delafossites, substituting Mg$^{2+}$ for Cr$^{3+}$ or Rh$^{3+}$, leads to interesting thermoelectric properties characteristic of p-type materials which can be described as doped semiconductors in the case of CuCr$_{1-x}$Mg$_x$O$_2$ but as metals in the case of heavily doped CuRh$_{1-x}$Mg$_x$O$_2$ ($x\simeq 0.1$).

Let us also remark that the band structure of delafossites entails Dirac
cones, especially the one of \curho{} \cite{Mai09b}. Since they are far away
from the Fermi energy they do not seem to have much influence on the physical
properties. Yet, the Dirac cones could be brought closer to the Fermi energy 
in a capacitor geometry
through the use of gate biases, see, {\em e.g.}, \cite{Thi06,Ste17}, to provide
us with the first oxide topological insulator.

The large splitting of the 3d (or 4d) $t_{2g}$ and $e_g$ orbitals at 
$E_F$ in \cucro{} (or \curho{}) which accounts for the physical properties 
of these delafossites is in marked contrast with the incomplete band filling 
of $d^9$ Pd$^+$ in \pdcoo{} or \pdcro{} with Pd 4d states almost exclusively 
responsible for electrical conductivity. More particularly, the 
$d_{xy,x^2-y^2}$ orbitals add the largest contribution to the DOS at $E_F$. 
It results a simple Fermi surface, characteristic of quasi-2D electronic
states. Thus, \pdcoo{} can be regarded as a textbook example of quasi-2D 
metals, which are also of high interest as paradigmatic candidates for 
angle-resolved photoemission spectroscopy line shape studies \cite{Cla96}. 
The stacking of Pd conducting layers alternating with a quasi-insulating 
CoO$_2$ layers leads to effects rarely seen in condensed matter. Although the 2D
character is reflected in the anisotropic resistivities, the thermal
conductivity $\kappa$ being dominated by the lattice contribution, $\kappa$ is
much less anisotropic. As a result, \pdcoo{} is a metal with anomalously large
$\kappa$ values as compared to the values reported for noble metals. The
thermoelectric power follows the prediction of the free-electron theory and,
for \pdcro{}, where the “insulating layer” contains a paramagnetic cations
(Cr$^{3+}$; S=3/2), the physics becomes more complex. For instance, the Nernst
coefficient is found to become large below $T_N$ with a very large sensitivity
to the magnetic field. It must be also emphasized that these metallic
delafossites belong also to the Weyl semimetals, a class of topological
materials. This hypothesis has been put forward to explain the negative
longitudinal magnetoresistance. Finally, growth of thicker crystals is needed
for testing the prediction of high Seebeck coefficient along the transverse
direction ($S\simeq $ \seebeck{200} along the c axis) together with a moderate
out-of-plane resistivity. However, the $\kappa$ measurement revealing very
large values even for the transverse direction are not in favor of any
applicability of these material in thermoelectricity based waste-heat
recovery.

Following the graphene original physics, the delafossites compounds, which 2D
structure can be described as a natural 1:1 epitaxy of a metal $d^9$ or
insulator $d^{10}$ metal layers with a MO$_2$ layer of the CdI$_2$-type, which
can be diamagnetic or paramagnetic at room temperature with possibly an
antiferromagnetic order at low $T$, offer a broad range of combinations to
generate new physical properties. In that respect, there remains room for the
chemists to produce materials of original compositions, to grow larger
crystals to characterize all anisotropic properties --– including the
thermopower and thermal conductivity --- and for the physicist, beyond all
measurements still to be done, to improve the modeling of these properties. 

Finally, it must be emphasized that the selected examples described in this
review paper represent only a few compounds among a very broad family of
delafossites. Looking beyond the oxides at compounds crystallizing in similar
structures, it must be also mentioned that three exists many layered
AMX$_2$ compounds for other chalcogen ions (X=S, Se, Te) for which the physics
is also very exciting as complex antiferromagnetism in AgCrS$_2$ and
AgCrSe$_2$ or the photovoltaic effects in CIGS
(CuGa$_{1-x}$In$_x$Se$_2$). However, to the best of our knowledge,
delafossite-like sulfides or selenides with metallicity as good as that of
\pdcoo{} keep on escaping synthesis. Again, this outlines the importance of oxides to generate unusual properties.

\section{Acknowledgements}
The authors are grateful for collaborations and discussions with numerous
colleagues, especially E.~Guilmeau, C.~Hicks, K.-H.~H\"ock, T.~Kopp,
S.~Kremer, D.~Ledue, C.~Martin, M.~Poienar, W.~C.~Sheets, and Ch.~Simon.
The authors acknowledge the financial support of the
French Agence Nationale de la Recherche (ANR), through
the program Investissements d’Avenir (ANR-10-LABX-09-01), LabEx EMC3, and the
Deutsche Forschungsgemeinschaft through SFB 484 and TRR 80.
Figs.\ \ref{figstruct} and \ref{dft:pdcoo2:frm} were generated using the 
XCrysDen software (Ref.\ \cite{Kok03}).

\end{document}